\documentclass[11pt,a4paper,oneside]{article}

\usepackage{mymacro}
\usepackage{verbatim}
\usepackage{tikz}
\usetikzlibrary{patterns,decorations.pathmorphing}
\usetikzlibrary{arrows.meta}
 \usetikzlibrary{shapes.geometric}
\definecolor{myRed}{RGB}{150,22,22}
\protected\edef\mathcal{\unexpanded\expandafter\expandafter\expandafter{  \csname mathcal \endcsname }}

\DeclareMathOperator{\dDisc}{dDisc}
\newcommand{\circdiag}[1]{\raisebox{0.15ex}{\tikz[baseline=(X.base)]{\node[draw,circle,inner sep=1.2pt] (X) {$#1$};}}}

\title{\textbf{Higher-Trace Operators and Cut Diagrammatics in the Conformal Block Expansion}}

\author{Mohammad Reza Khansari}

\affiliation{SISSA, Via Bonomea 265, 34136 Trieste, Italy}
\affiliation{INFN, Sezione di Trieste, Via Valerio 2, 34127 Trieste, Italy}

\makeatletter
\renewcommand{\@email}[1]{#1}
\makeatother

\emailAdd{
            \href{mailto:mkhansar@sissa.it}{\tt mkhansar@sissa.it}
}

\abstract{
We study four-point functions of identical scalar operators in conformal field theories with AdS duals in large-$N$ expansion. We analyze the appearance of higher-trace operators in theories dual to bulk $\phi^3$ and $\phi^4$ interactions, focusing on how these operators are required by crossing symmetry. We compute part of the OPE data associated with these operators. We also introduce a diagrammatic framework for organizing the different terms in the conformal block expansion within the large-$N$ expansion. This framework refines the use of crossing symmetry by allowing it to be applied to individual diagrammatic topologies, rather than only to the full correlator. It further separates different contributions to the OPE data by associating them with different cut diagrams. In this language, the emergence of higher-trace operators and their relation to lower-trace OPE data become more manifest.}

\begin{document}

\maketitle

\section{Introduction}

Scattering amplitudes in curved spacetimes are difficult to define in the same way as in flat space. In Anti-de Sitter (AdS) space, the absence of standard asymptotic states prevents a direct definition of the usual S-matrix. Nevertheless, the AdS/CFT correspondence provides a precise framework in which questions about bulk scattering can be translated into questions about boundary correlation functions \cite{Maldacena:1997re,Witten:1998qj,Gubser:1998bc,Heemskerk:2009pn,Penedones:2010ue,Fitzpatrick:2011ia,Fitzpatrick:2011hu,Fitzpatrick:2011dm}.

In this paper we study the four-point function of a scalar operator $\mathcal O$ in Conformal Field Theories (CFTs) with weakly coupled AdS duals in large-$N$ expansion. The operator $\mathcal O$ is dual to a bulk scalar field $\phi$. $N$ denotes the large parameter controlling the number of degrees of freedom of the CFT. The large-$N$ expansion of the CFT correlator is dual to the loop expansion of Witten diagrams in AdS.

The conformal block decomposition organizes each order in this expansion in terms of the CFT operators exchanged in a chosen Operator Product Expansion (OPE) channel. Crossing symmetry then imposes non-trivial constraints on their OPE data. This is the basic logic of the conformal bootstrap program~\cite{El-Showk:2011yvt,Fitzpatrick:2012yx,Komargodski:2012ek,Fitzpatrick:2014vua,Alday:2014tsa,Kaviraj:2015cxa,Kaviraj:2015xsa,Alday:2015eya,Alday:2016njk,Aharony:2016dwx,Alday:2017gde,Poland:2018epd,Albayrak:2019gnz,Bissi:2022mrs,Huang:2023oxf,BubblesAdS} in large-$N$ expansion, and in holographic theories it gives a way of reconstructing bulk dynamics from CFT data.

At leading order in $1/N$, crossing symmetry reconstructs tree-level exchange amplitudes from the appearance of single-trace operators in the $\mathcal O\times\mathcal O$ OPE \cite{Alday:2017gde}. At one loop,  in the scalar theory dual to a bulk $\phi^4$ interaction, the double-trace OPE data at order $1/N^4$ can be reconstructed by crossing symmetry from tree-level input \cite{Aharony:2016dwx}.\footnote{ Related one-loop double-trace data in $\phi^4$ theory have also been computed directly in  AdS$_4$~\cite{Bertan:2018afl} and AdS$_3$ \cite{Carmi:2026spv}. See also~\cite{Akhmedov:2018lkp,Cacciatori:2024zbe,Carmi:2019ocp,Carmi:2021dsn,Carmi:2024tzj,Bertan:2018khc,Xiao:2026prw,Albayrak:2020bso} for more works on loop physics in AdS using direct AdS computations.}

This one-loop reconstruction was extended to higher-loop bubble diagrams in~\cite{BubblesAdS}. Using the Lorentzian inversion formula \cite{Caron-Huot:2017vep}, which extracts CFT data from crossed-channel double-discontinuities, it was shown that at any order in $1/N$ there is a universal part of the double-trace OPE data completely fixed by tree-level data. In particular, at order $N^{-2k}$ for $k\geq2$ this contribution to the anomalous dimensions behaves at large spin $\ell$ as
\begin{equation} \gamma_{\ell} \sim \frac{\log^{k-2}\ell}{\ell^{2\Delta_\mathcal{O}}},
\end{equation}
where $\Delta_\CO$ is the scaling dimension of the boundary operator $\CO$. It was further shown in~\cite{BubblesAdS} that this contribution admits a diagrammatic interpretation in terms of maximal cuts of $(k-1)$-loop bubble Witten diagrams.\footnote{A similar study in the case of $\mathcal N=4$ SYM  was carried out in \cite{Bissi:2020wtv,Bissi:2020woe}.} Moreover, in Mellin space these maximal cuts were shown to reduce, in the flat-space limit, to the familiar consecutive unitarity cuts of flat-space bubble diagrams. This will serve as motivation for the diagrammatic notation we later introduce.

One aim of the present work is to understand what additional structures are generated in the conformal block expansion once infinite spin sums of double-trace data are treated carefully. A key point is that such sums can produce behaviours in small cross ratio limit which is not visible at the level of individual blocks. After applying crossing symmetry, these terms can have the power law structure expected from the exchange of higher-trace operators. In this way, higher trace operators appear as a necessary part of the crossing-symmetric completion of the lower-trace data.

We analyze this mechanism in theories dual to bulk $\phi^4$ and $\phi^3$ interactions. For the $\phi^4$ case, at order $N^{-8}$ the crossed image of double-trace spin sums gives rise to quadruple-trace operators. In the $\phi^3$ case, the same mechanism generates higher-trace sectors, with the $k-$trace operators appearing at order $N^{-2k}$ for $k\geq3$. We compute part of the OPE data associated with these operators, either in closed form in special cases or in a large-spin expansion. This connects to the problem of understanding multi-particle states in AdS from CFT data \cite{Fardelli:2025eun,Kravchuk:2024wmv}.

We also introduce a diagrammatic representation for organizing products of OPE data and the corresponding terms in the conformal block expansion. The diagrams in this representation contain cuts, but these cuts need not be maximal. They should therefore be distinguished from the maximal cuts appearing in~\cite{BubblesAdS}, whose flat-space limit was shown to reproduce ordinary unitarity cuts. In the more general setting considered here, we will not attempt to make an explicit relation between our cut diagrams and unitarity cuts. However, this remains as something to be explored in the future. 

The purpose of the diagrammatic representation is to provide a bookkeeping device for CFT data, and to give both OPE data and terms in the conformal block expansion a more clear amplitude-like interpretation. In this language, crossing symmetry can be applied to subsets of terms associated with a given diagrammatic topology, rather than only to the full correlator. This is particularly useful for the higher-trace terms discussed above, whose origin can be traced to specific crossed-channel spin sums.

At the level of OPE data, the same representation also suggests a natural separation between different channel contributions. For example as we will see, for scalar exchange, the asymptotic large-spin solution is naturally associated with the $t$-channel and $u$-channel exchange diagram, while the finite-spin part is associated with the $s$-channel exchange diagram. We test these assignments by computing one-loop Mellin amplitudes in scalar $\phi^3+\phi^4$ theory.

The paper is organized as follows. In Section~\ref{sec1} we introduce the general tools and method. We also review the double-trace data associated with bubble diagrams. In Section~\ref{sec3} we study the appearance of higher-trace operators in theories dual to $\phi^4$ and $\phi^3$ interactions, and compute part of the associated OPE data. In Section~\ref{sec4} we introduce the diagrammatic representation for products of OPE data and for the corresponding terms in the conformal block expansion.  Appendix~\ref{appxA} contains technical details on the spin sums used in the main text.  Appendix~\ref{app2} reviews Mellin-space conventions and the flat-space limit.

\section{Setup and double-trace data}\label{sec1}
We consider the four-point function of identical scalar primary operators $\mathcal{O}$ of scaling dimension $\Delta_\mathcal{O}$ in a four-dimensional CFT. Conformal invariance fixes its spacetime dependence to be
\begin{equation}
\langle \mathcal{O}(x_1) \mathcal{O}(x_2) \mathcal{O}(x_3) \mathcal{O}(x_4)\rangle=\frac{1}{(x_{12}^2x_{34}^2)^{\Delta_\mathcal{O}}} \mathcal{G}(u,v),
\end{equation}
where $x_{ij}^2=(x_i-x_j)^2$, and
\begin{equation}
u=\frac{x_{12}^2x_{34}^2}{x_{13}^2x_{24}^2} \equiv z\bar z,\qquad v=\frac{x_{14}^2x_{23}^2}{x_{13}^2x_{24}^2} \equiv (1-z)(1-\bar z).
\end{equation}
For identical external operators, crossing symmetry under the exchange
$1\leftrightarrow 3$ gives
\begin{equation}\label{crossing}
\mathcal{G}(u,v) = \left(\frac{u}{v}\right)^{\Delta_\mathcal{O}} \mathcal{G}(v,u).
\end{equation}

We decompose the correlator into four-dimensional scalar conformal blocks as~\cite{Dolan:2003hv}
\begin{equation}\label{blockexp}
\mathcal{G}(u,v) = \sum_{\Delta,\ell} a_{\Delta,\ell}\, g_{\Delta,\ell}(u,v), \qquad a_{\Delta,\ell} = f_{\mathcal{O}\mathcal{O}\mathcal{O}_{\Delta,\ell}}^2 .
\end{equation}
with
\begin{equation}\label{DOblock}
g_{\Delta,\ell}(z,\bar z) =\frac{z\bar z}{z-\bar z} \left[ k_{\frac{\Delta+\ell}{2}}(z) k_{\frac{\Delta-\ell-2}{2}}(\bar z)-(z\leftrightarrow \bar z)\right],
\end{equation}
\begin{equation}\label{kblock}
k_{\beta}(x)= x^\beta\, F_{\beta}(x), \qquad F_{\beta}(x)={}_2F_1(\beta,\beta,2\beta;x).
\end{equation}
The sum runs over primary operators appearing in the $\mathcal{O}\times\mathcal{O}$ OPE. Since the external operators are identical scalars, only even spins $\ell$ appear.

It will be useful to factor out the leading power of $u$ from a block and define a reduced block as
\begin{equation}
  g_{\Delta,\ell}  =u^{\frac{\Delta-\ell}{2}}\tilde{g}_{\Delta,\ell}
\end{equation}

We consider a CFT at large-$N$ expansion. At leading order in the expansion, the four-point function is that of generalized free field (GFF) theory. Namely,
\begin{equation}\label{GFF}
\mathcal{G}^{(0)}(u,v) =1+u^{\Delta_\mathcal{O}}+\left(\frac{u}{v}\right)^{\Delta_\mathcal{O}} .
\end{equation}
The conformal block decomposition of this correlator contains the identity and the tower of double-trace operators of the schematic form:
\begin{equation}
[\mathcal{O}\mathcal{O}]_{n,\ell}\sim\mathcal{O}\,\square^n\partial_{\mu_1}\cdots\partial_{\mu_\ell}\mathcal{O},
\end{equation}
with scaling dimensions $\Delta_{n,\ell}^{(0)}= 2\Delta_\mathcal{O}+2n+\ell ,$ and corresponding OPE coefficients~\cite{Heemskerk:2009pn}
\begin{equation}\label{a0}
a_{n,\ell}^{(0)}=\frac{2(\ell+1)(2\Delta_\mathcal{O}+2n+\ell-2)}{(\Delta_\mathcal{O}-1)^2}\,C_n^{\Delta_\mathcal{O}-1}\,C_{n+\ell+1}^{\Delta_\mathcal{O}-1},
\end{equation}
where
\begin{equation}
C_n^\Delta=\frac{\Gamma^2(\Delta+n)\Gamma(2\Delta+n-1)}{\Gamma(n+1)\Gamma^2(\Delta)\Gamma(2\Delta+2n-1)}.
\end{equation}

Beyond the leading order we expand the correlator as
\begin{equation}\label{Gexp}
\mathcal{G}(u,v)=\mathcal{G}^{(0)}(u,v)+\frac{1}{N^2}\mathcal{G}^{(1)}(u,v)+\frac{1}{N^4}\mathcal{G}^{(2)}(u,v)+\cdots.
\end{equation}
The OPE data of double-trace operators admit a corresponding expansion,
\begin{align}
\Delta_{n,\ell}&=2\Delta_\mathcal{O}+2n+\ell+\frac{\gamma_{n,\ell}^{(1)}}{N^2}+\frac{\gamma_{n,\ell}^{(2)}}{N^4}+\frac{\gamma_{n,\ell}^{(3)}}{N^6}+\cdots ,\\a_{n,\ell}&=a_{n,\ell}^{(0)}+\frac{a_{n,\ell}^{(1)}}{N^2}+\frac{a_{n,\ell}^{(2)}}{N^4}+\cdots .
\end{align}

To find the OPE data at each order of the expansion, the main tool we use  is the Lorentzian inversion formula~\cite{Caron-Huot:2017vep}. For identical scalar operators in four dimensions we write
\begin{equation}\label{LIF}
c(\Delta,\ell)=\frac{1+(-1)^\ell}{4}\,\kappa_{\frac{\Delta+\ell}{2}}\int_0^1 dz\,d\bar z\,\frac{(z-\bar z)^2}{z^4\bar z^4}\,g_{\ell+3,\Delta-3}(z,\bar z)\,\dDisc[\mathcal{G}(z,\bar z)],
\end{equation}
with
$\kappa_\alpha=\frac{\Gamma^4(\alpha)}{2\pi^2\Gamma(2\alpha-1)\Gamma(2\alpha)}.$ The double discontinuity is defined by
\begin{equation}
\dDisc[\mathcal{G}(z,\bar z)]= \mathcal{G}(z,\bar z)-\frac12\mathcal{G}^{\circlearrowright}(z,\bar z)-\frac12 \mathcal{G}^{\circlearrowleft}(z,\bar z),
\end{equation}
where the two analytic continuations are taken around $\bar z=1$.

The poles and residues of the function $c(\Delta,\ell)$ encode the spectrum and OPE coefficients of the exchanged operators. Near the pole associated with an operator of dimension $\Delta_\ast$ and spin $\ell$, we have
\begin{equation}\label{pole}
c(\Delta,\ell)\sim\frac{a_{\Delta_\ast,\ell}}{\Delta_\ast-\Delta}.
\end{equation}

\paragraph{Generic exchange}

Let $U$ denote a generic primary operator in the $\mathcal{O}\times\mathcal{O}$ OPE of dimension $\Delta_U$, spin $\ell_U$, twist $\tau_U=\Delta_U-\ell_U$, and squared OPE coefficient $a_{\Delta_U,\ell_U}$. Its direct-channel contribution to the correlator can be written as
\begin{equation}
\mathcal{G}(u,v)\supset a_{\Delta_U,\ell_U}\,u^{\tau_U/2} \,\tilde g_U(u,v),
\end{equation}
where $\tilde g_U$ denotes the corresponding reduced block. By crossing symmetry, this contribution implies a term in the  crossed channel
\begin{equation}\label{cr}
\mathcal{G}(u,v)
\supset a_{\Delta_U,\ell_U}\,u^{\Delta_\mathcal{O}} \,v^{\tau_U/2-\Delta_\mathcal{O}} \,\tilde g_U(v,u).
\end{equation}

Inserting~\eqref{cr} into the Lorentzian inversion formula and taking the same limits of \cite{BubblesAdS} gives the contribution of $U$ to the anomalous dimensions of double trace operators at large spin expansion.  If $U$ first enters the OPE with coefficient of order $N^{-p}$, then its contribution to the correlator starts at order $N^{-2p}$. For leading twist, the large-spin anomalous dimension takes the form
\begin{equation}\label{gaml0}
N^{2p}\gamma_{0,\ell} = \frac{\mathcal{A}(\Delta_\mathcal{O},\ell_U,\tau_U)}{J^{\tau_U}} + \frac{\mathcal{B}(\Delta_\mathcal{O},\ell_U,\tau_U)}{J^{\tau_U+2}} +\mathcal{O}\!\left(\frac{1}{J^{\tau_U+4}} \right).
\end{equation}
where $J^2=(\ell+\Delta_\mathcal{O})(\ell+\Delta_\mathcal{O}-1)$ is the conformal spin for the leading twist double trace operator.

The coefficients $\mathcal{A}$ and $\mathcal{B}$ are given by
\begin{equation}\label{Acoef}
\mathcal{A} =-2a_{\Delta_U,\ell_U}\frac{\Gamma(\Delta_\mathcal{O})^2} {\Gamma\!\left(\Delta_\mathcal{O}-\frac{\tau_U}{2}\right)^2} \frac{\Gamma(2\ell_U+\tau_U)}{\Gamma\!\left(\ell_U+\frac{\tau_U}{2}\right)^2} \left[ 1- \frac{\Gamma(\tau_U-2)} {\Gamma(-\ell_U)\Gamma(\ell_U+\tau_U-1)} \right],
\end{equation}
\begin{equation}\label{Bcoef}
\mathcal{B}=\frac{a_{\Delta_U,\ell_U}}{24}\frac{\Gamma(\Delta_\mathcal{O})^2}{\Gamma\!\left(\Delta_\mathcal{O}-\frac{\tau_U}{2}\right)^2} \frac{\Gamma(2\ell_U+\tau_U)}{\Gamma\!\left(\ell_U+\frac{\tau_U}{2}\right)^2}\left[\mathcal{P}(\Delta_\mathcal{O},\tau_U)+ \frac{ \mathcal{R}(\Delta_\mathcal{O},\ell_U,\tau_U) \Gamma(\tau_U-2) }{\ell_U\Gamma(-\ell_U)\Gamma(\ell_U+\tau_U)}\right],
\end{equation}
where
\begin{equation}
\mathcal{P}(\Delta_\mathcal{O},\tau_U)=\tau_U^3-4\tau_U\Big(1+3(\Delta_\mathcal{O}-2)\Delta_\mathcal{O}\Big)-24\left(1-\Delta_\mathcal{O}+\frac{\tau_U}{2}\right)^2 ,
\end{equation}
and
\begin{equation}
\begin{aligned}
\mathcal{R}(\Delta_\mathcal{O},\ell_U,\tau_U)
&= 12\Delta_\mathcal{O}^2 \left[ 4\ell_U^2-(\tau_U-2)^2+4\ell_U(\tau_U-1)\right]-3(\tau_U^2-4)^2\\&\quad -4\ell_U^2(-12-10\tau_U+\tau_U^3)-4\ell_U(\tau_U-1)(-12-10\tau_U+\tau_U^3)\\&\quad +12\Delta_\mathcal{O}\Big[ (\tau_U-2)^2(\tau_U+2)+\ell_U^2(\tau_U^2-4\tau_U-8)\\&\hspace{3.8cm} +\ell_U(\tau_U-1)(\tau_U^2-4\tau_U-8) \Big].
\end{aligned}
\end{equation}
These formulae will be used below both for single-trace and double-trace exchanges. 

For the case of single trace exchange, as a simple example, suppose that $U=\mathcal{O}$. Then $p=1$, $\tau_U=\Delta_\mathcal{O}$, and $\ell_U=0$. Inserting these data in \eqref{gaml0} gives
\begin{equation}\label{g225}
\begin{aligned}
\gamma^{(1),\mathrm{as}}_{0,\ell}
&= -\frac{ 2a_{\Delta_\mathcal{O},0} \Gamma(\Delta_\mathcal{O})^3}{\Gamma\!\left(\frac{\Delta_\mathcal{O}}{2}\right)^4}\frac{1}{J^{\Delta_\mathcal{O}}}
\\
&\quad -\frac{ a_{\Delta_\mathcal{O},0} \Gamma(\Delta_\mathcal{O})^3}{24\,\Gamma\!\left(\frac{\Delta_\mathcal{O}}{2}\right)^4}\frac{\Delta_\mathcal{O}(\Delta_\mathcal{O}-2)\left(11\Delta_\mathcal{O}^2-10\Delta_\mathcal{O}-4\right)}{\Delta_\mathcal{O}-1}\frac{1}{J^{\Delta_\mathcal{O}+2}}+\mathcal{O}\!\left(\frac{1}{J^{\Delta_\mathcal{O}+4}}\right).
\end{aligned}
\end{equation}
Here $a_{\Delta_\mathcal{O},0}=f_{\mathcal{O}\mathcal{O}\mathcal{O}}^2$, and the superscript ``$\mathrm{as}$'' denotes the asymptotic large-spin solution. This solution has infinite support in spin. Crossing symmetry also requires an additional finite-spin contribution~\cite{Alday:2017gde}. In Section~\ref{sec4}, we will distinguish these asymptotic and finite pieces diagrammatically.

\subsection{Bubble data}

We now turn to the case in which the exchanged operator $U$ itself belongs to the double-trace tower. In this case, one should first observe that the result \eqref{gaml0} vanishes whenever $\tau_U-2\Delta_\mathcal{O}\in 2\mathbb{Z}_{\geq0}$, which precisely corresponds to the bare twists of double trace operators. The non-trivial contribution appears only after including anomalous dimensions and in general after summing over the relevant tower of exchanged operators.

For a finite sum of exchanged operators one may freely interchange the double discontinuity in the inversion formula with the sum. For an infinite tower this is not automatic. Namely:
\begin{equation} \label{ineq}
\dDisc\!\left[ \sum_{n,\ell} h_{n,\ell} \right]\neq\sum_{n,\ell}\dDisc[h_{n,\ell}], \qquad \sum_{n,\ell} h_{n,\ell} =\mathcal{G}(u,v).
\end{equation}
Infinite spin sums can generate singularities which are invisible term by term. This mechanism will be central in Section~\ref{sec3}, and we discuss some general cases in the appendix \ref{appxA}.

A sufficient condition to make the inequality of \eqref{ineq} into equality is that the function $h_{n,\ell}$ has finite support in spin. This will be realized in the example of quartic interaction without derivative in the bulk. In this case, the tree-level OPE data have support only at spin zero \cite{Heemskerk:2009pn},
\begin{equation}
\gamma^{(1)}_{n,\ell}=\frac{(2\Delta_\mathcal{O}-1)(n+1)(2\Delta_\mathcal{O}+n-3)(\Delta_\mathcal{O}+n-1)}{(\Delta_\mathcal{O}-1)(2\Delta_\mathcal{O}+2n-3)(2\Delta_\mathcal{O}+2n-1)}\,\delta_{\ell,0},
\end{equation}
\begin{equation}\label{phi4a1}
a^{(1)}_{n,0}=\frac12\partial_n\left(a^{(0)}_{n,0}\gamma^{(1)}_{n,0}\right).
\end{equation}
Therefore, because of the finite support in spin,  one may treat each exchanged operator independently and  afterwards sum over the level $n$.

We introduce the shorthand
\begin{equation}
\epsilon_n \equiv \frac{\gamma^{(1)}_{n,0}}{N^2},\qquad\alpha_n \equiv a^{(0)}_{n,0}+\frac{a^{(1)}_{n,0}}{N^2}.
\end{equation}

We define an effective scalar exchange with
\begin{equation}\label{tudouble}
\tau_U=2\Delta_\mathcal{O}+2n+\epsilon_n,\qquad\ell_U=0,\qquad a_{\Delta_U,0}=\alpha_n .
\end{equation}
This leads to
\begin{equation}
\widehat{\gamma}_{0,\ell}=\sum_{n,m\geq0}\frac{f_{nm}}{J^{2\Delta_\mathcal{O}+2n+2m+\epsilon_n}}.
\end{equation}
The hat indicates that we keep only the contribution determined by tree-level OPE data.

This effective description was briefly discussed in~\cite{BubblesAdS} and admits a simple bulk interpretation. At one loop, the spectral decomposition of the product of the two internal propagators expresses the bubble Witten diagram as a sum of tree-level scalar exchanges with dimensions $2\Delta_\mathcal{O}+2n$~\cite{Fitzpatrick:2011dm}. In~\cite{BubblesAdS}, this picture was extended to higher orders by shifting these dimensions by the tree-level anomalous dimensions, as in~\eqref{tudouble}. As will be seen from the explicit coefficients below, $f_{nm}$ start at order $\epsilon_n^2$. Consequently, expanding
\begin{equation}
    J^{-\epsilon_n}=e^{-\epsilon_n\log J}
\end{equation}
generates a term proportional to $\log^{k-2}J$ at order $N^{-2k}$. This is exactly the leading logarithmic contribution at that order determined by tree-level OPE data \cite{BubblesAdS}. 

This term-by-term effective description is valid here because the tree-level data have finite support in spin, so that the two sides of~\eqref{ineq} are equal. It cannot, in general, be extended simply by including $\gamma^{(2)}_{n,\ell}/N^4$ and higher corrections in the effective dimensions. Indeed, $\gamma^{(2)}_{n,\ell}$ has infinite support in spin, and the interchange of the spin sum with the double discontinuity is then no longer justified.

The analysis of~\cite{BubblesAdS} focused on $\Delta_\mathcal{O}=2$. Here we extend this description to generic $\Delta_\mathcal{O}$ by reporting the first few coefficients $f_{nm}$. Using~\eqref{Acoef} and~\eqref{Bcoef}, we find
\begin{equation}\label{f00i}
f_{00} =-2\alpha_0\frac{\Gamma(\Delta_\mathcal{O})^2\Gamma(2\Delta_\mathcal{O}+\epsilon_0)}{\Gamma\!\left(-\frac{\epsilon_0}{2}\right)^2\Gamma\!\left(\Delta_\mathcal{O}+\frac{\epsilon_0}{2}\right)^2},
\end{equation}

\begin{equation}\label{f10i}
f_{10}=-2\alpha_1\frac{\Gamma(\Delta_\mathcal{O})^2\Gamma(2\Delta_\mathcal{O}+2+\epsilon_1)}{\Gamma\!\left(-1-\frac{\epsilon_1}{2}\right)^2\Gamma\!\left(\Delta_\mathcal{O}+1+\frac{\epsilon_1}{2}\right)^2},
\end{equation}

\begin{equation}\label{f01i}
f_{01}=-\frac{\mathcal{S}_0}{48}f_{00},
\end{equation}
where
\begin{equation}
\mathcal{S}_0=\frac{x_0\left[x_0^3-4x_0^2-12\Delta_\mathcal{O}^2x_0+36\Delta_\mathcal{O}x_0-16x_0-8\right]}{x_0-1},\qquad x_0= 2\Delta_\mathcal{O} +\epsilon_0.
\end{equation}

The same method can be used to obtain arbitrarily high orders in the large-spin expansion.

Expanding these expressions at large $N$ gives loop-level double-trace data. At one loop, namely at order $N^{-4}$, we find
\begin{equation}\label{gamma2}
\gamma^{(2)}_{0,\ell}= -\frac{\Gamma(2\Delta_\mathcal{O})}{J^{2\Delta_\mathcal{O}}}+ \frac{ \Delta_\mathcal{O} \left[1+\Delta_\mathcal{O}\left(8-\Delta_\mathcal{O}(7+2\Delta_\mathcal{O})\right)\right]\Gamma(2\Delta_\mathcal{O})}{3(1+2\Delta_\mathcal{O})J^{2\Delta_\mathcal{O}+2}}+
\mathcal{O}\!\left(\frac{1}{J^{2\Delta_\mathcal{O}+4}}\right).
\end{equation}
At two loops, this will fix the $\log J$ coefficient of the anomalous dimension. Namely: 
\begin{equation}\label{gamma3log}
\gamma^{(3)}_{0,\ell}\Big|_{\log J}=\frac{\Gamma(2\Delta_\mathcal{O})}{J^{2\Delta_\mathcal{O}}}+\frac{ (\Delta_\mathcal{O}-1)\Delta_\mathcal{O}\left[1+\Delta_\mathcal{O}\left(11+4\Delta_\mathcal{O}(11+\Delta_\mathcal{O})\right)\right]\Gamma(2\Delta_\mathcal{O})}{3(1+2\Delta_\mathcal{O})^2J^{2\Delta_\mathcal{O}+2}}+\mathcal{O}\!\left(\frac{1}{J^{2\Delta_\mathcal{O}+4}}\right).
\end{equation}
More generally, the part of the higher-loop OPE data fixed by tree-level input is the leading logarithmic term in the large-spin expansion~\cite{BubblesAdS}. Thus, at order $N^{-2k}$, the tree-level data contribution to the double-trace anomalous dimension completely fixes
\begin{equation}
\gamma^{(k)}_{n,\ell}\Big|_{\log^{k-2}J}.
\end{equation}

\subsubsection{Subleading twist}

So far we have focused mostly on the leading-twist double-trace family $[\mathcal{O}\mathcal{O}]_{0,\ell}$. In the next sections we will also need the large-spin behavior of the full tower $[\mathcal{O}\mathcal{O}]_{n,\ell}$. The analysis of the previous subsection can be systematically extended to subleading twists, but for our purposes we only need the leading large-spin behavior at fixed $n$. We therefore briefly recall the standard large-spin result of~\cite{Kaviraj:2015cxa}.

For identical external scalar operators of dimension $\Delta_\CO$, the anomalous dimension of the double-trace operator $[\CO\CO]_{n,\ell}$ induced by the exchange of an operator with twist $\tau_U$ and spin $\ell_U$ has the leading large-spin form
\begin{equation}
    \gamma_{n,\ell} =   \frac{\gamma_n}{J^{\tau_U}} +  \mathcal{O}\!\left(\frac{1}{J^{\tau_U+2}}\right),
\end{equation}
where for the family $[\CO\CO]_{n,\ell}$ we use the conformal spin defined as $ J^2  = (\ell+\Delta_\CO+n)(\ell+\Delta_\CO+n-1).$ The coefficient $\gamma_n$ is determined by a finite sum
\begin{equation} \label{subt}
    \gamma_n= \sum_{m=0}^{n} c_{n,m} L_m,
\end{equation}
where
\begin{equation}
    c_{n,m}= \frac{1}{8}\left[\frac{\Gamma(\Delta_\CO)}{\Gamma(\Delta_\CO+m-1)}\right]^2 (2\Delta_\CO+n-3)_m(-1)^{n+m} \frac{n!}{(n-m)!}\left[\frac{\Gamma(\Delta_\CO-1)}{\Gamma\!\left(\Delta_\CO-\frac{\tau_U}{2}\right)}\right]^2 ,
\end{equation}
and
\begin{equation}
    L_m =-16\, a_{\Delta_U,\ell_U} \frac{ \Gamma(2\ell_U+\tau_U)\Gamma\!\left(m+\ell_U+\frac{\tau_U}{2}\right)^2 }{ \Gamma(m+1)^2\Gamma\!\left(\ell_U+\frac{\tau_U}{2}\right)^4}\,{}_3F_2\!\left[
    \begin{matrix}
    -m,\,-m,\,-1-\ell_U+\Delta_\CO-\frac{\tau_U}{2}\\ 1-m-\ell_U-\frac{\tau_U}{2},\, 1-m-\ell_U-\frac{\tau_U}{2}
    \end{matrix}
    ;1 \right].
\end{equation}
Here we remind that $a_{\Delta_U,\ell_U}$ denotes the squared OPE coefficient of the exchanged operator.

As an example, at $n=0$ only the $m=0$ term contributes in \eqref{subt}, and one finds
\begin{equation}
    \gamma_0 =-2a_{\Delta_U,\ell_U} \frac{\Gamma(\Delta_\CO)^2} {\Gamma\!\left(\Delta_\CO-\frac{\tau_U}{2}\right)^2} \frac{\Gamma(2\ell_U+\tau_U)} {\Gamma\!\left(\ell_U+\frac{\tau_U}{2}\right)^2}.
\end{equation}

Let us now specialize this formula to the case of $\phi^4$ interaction in the bulk. As discussed earlier, in order to obtain the effect of tree-level data at all loops, we can consider an effective exchange with scaling dimension and OPE coefficient given in \eqref{tudouble}. Inserting this into \eqref{subt} and expanding in $1/N$, one finds at order $N^{-4}$:
\begin{equation}\label{gam242}
    \gamma^{(2)}_{n,\ell}  =\frac{\zeta_n}{J^{2\Delta_\CO}}+\mathcal{O}\!\left(\frac{1}{J^{2\Delta_\CO+2}}\right),
\end{equation}
where the coefficient $\zeta_n$ can be found in closed form
\begin{equation}\label{zetan}
    \zeta_n= 2(1-2\Delta_\CO) \left[2\Delta_\CO^2 +(4n-5)\Delta_\CO +2n^2-6n+3 \right] \frac{ \Gamma(2\Delta_\CO+n-3) }{ \Gamma(n+1) } .
\end{equation}

At order $N^{-4}$, the result is completely fixed by this effective tree-level exchange. At order $N^{-6}$, the tree-level data fix only the coefficient of $\log J$ in the anomalous dimension. Therefore we find
\begin{equation}\label{gam3lll}
    \gamma^{(3)}_{n,\ell}\Big|_{\log J} = -\frac{\zeta_n}{J^{2\Delta_\CO}}+ \mathcal{O}\!\left(\frac{1}{J^{2\Delta_\CO+2}}\right).
\end{equation}

In the following section, these large-spin data will be inserted into spin sums. 

\section{Higher-trace operators}\label{sec3}

At higher orders in the $1/N$ expansion multi-trace operators appear in the $\mathcal{O}\times\mathcal{O}$ OPE. Thus, beyond the double-trace sector, one must also keep track of the possible exchange of higher-trace operators.

A simplification can be obtained by imposing a $\mathbb{Z}_2$ symmetry under $\mathcal{O}\to-\mathcal{O}$. This forbids operators built from an odd number of $\mathcal{O}$'s in the $\mathcal{O}\times\mathcal{O}$ OPE. This is the situation relevant for a bulk theory with a $\phi^4$ interaction, and is the setup used in~\cite{BubblesAdS}. Even in this restricted setup, higher-trace operators with an even number of single trace constituents appear at subleading orders in the large-$N$ expansion. However, these operators do not affect the coefficient of the leading $\log J$ in the large-spin expansion of the double-trace OPE data studied in
\cite{BubblesAdS}. 

In this section, we are interested in the OPE data of the higher-trace operators themselves. We first analyze the appearance of quadruple-trace operators in the $\phi^4$ theory. We then discuss the $\phi^3$ theory, where the $\mathbb{Z}_2$ symmetry is absent and $k$-trace operators appear at order $N^{-2k}$ for integer $k\geq3$.

\subsection{$\phi^4$ theory}

In the $\mathbb{Z}_2$-symmetric case, the  OPE takes the schematic form
\begin{equation}\label{opephi4}
\mathcal{O}\times \mathcal{O} = \mathds{1} + [\mathcal{O}\mathcal{O}]_{n,\ell}+ \frac{1}{N^4}[\mathcal{O}^4]_{q,\ell,I} +\cdots .
\end{equation}
Here $[\mathcal{O}^4]_{q,\ell,I}$ denotes a quadruple-trace primary built from four copies of $\mathcal O$. The label $q$ denotes its twist level, and $I$ denotes a possible degeneracy label. Unlike in the double-trace sector, the labels $q$ and $\ell$ need not specify a unique quadruple-trace primary. At fixed total twist and spin there can be several independent quadruple-trace primaries, corresponding to different ways of distributing derivatives among the four single-trace constituents.

The coupling of quadruple-trace operators to the external pair is suppressed by $N^{-4}$, provided the quadruple-trace operators are orthogonal to lower-trace families. Therefore their squared OPE coefficients first contribute to the connected four-point function at order $N^{-8}$.

We expand the scaling dimension of quadruple-trace operators in large $N$  as
\begin{equation}
\Delta_{[\mathcal{O}^4]_{q,\ell,I}} = 4\Delta_\mathcal{O} +2q+\ell+\frac{\delta^{(1)}_{q,\ell,I}}{N^2}+\cdots ,
\end{equation}
and the corresponding squared OPE coefficients as
\begin{equation}
a_{[\mathcal{O}^4]_{q,\ell,I}}=\frac{b^{(0)}_{q,\ell,I}}{N^8}+\frac{b^{(1)}_{q,\ell,I}}{N^{10}}+\cdots .
\end{equation}
After summing over the degeneracy label, we define
\begin{equation}
    b^{(0)}_{q,\ell} \equiv \sum_I b^{(0)}_{q,\ell,I},
\end{equation}

\begin{equation}
    b^{(0)}_{q,\ell}\delta^{(1)}_{q,\ell}\equiv \sum_I b^{(0)}_{q,\ell,I}\delta^{(1)}_{q,\ell,I}.
\end{equation}

So the contribution of quadruple-trace operators to the correlator at order $N^{-8}$ takes the form 
\begin{equation}\label{gquad}
\mathcal{G}^{(4)}(u,v)\supset \sum_{q,\ell}b^{(0)}_{q,\ell}\,u^{2\Delta_\mathcal{O}+q}\,\tilde{g}^{(4)}_{q,\ell}(u,v),
\end{equation}
where
\begin{equation}
g_{4\Delta_\mathcal{O}+2q+\ell,\ell}(u,v)= u^{2\Delta_\mathcal{O}+q} \tilde{g}^{(4)}_{q,\ell}(u,v).
\end{equation}

We now consider the contribution to the same order coming from the double-trace sector. Among the various terms, there is in particular
\begin{equation}\label{dtlog}
\mathcal{G}^{(4)}(u,v)\supset\sum_{n,\ell}\frac{1}{8}\,a^{(0)}_{n,\ell}\,u^{\Delta_\mathcal{O}+n} \left(\gamma^{(2)}_{n,\ell}\right)^2\log^2 u\, \tilde{g}_{n,\ell}(u,v).
\end{equation}
As we reviewed in Section~\ref{sec1}, we can write
\begin{equation}\label{gamm22}
    \left(\gamma^{(2)}_{n,\ell}\right)^2   =\sum_{\rho\geq 0}\frac{c_{n,\rho}}{J^{4\Delta_\mathcal{O}+2\rho}} .
\end{equation}

Inserting this into~\eqref{dtlog}, each term labeled by $(n,\rho)$ gives a spin sum of the form $H_n^{(2\Delta_\CO+\rho)}(z,\bar z)$, where $H_n^{(m)}$ denotes the weighted sum over double-trace conformal blocks at fixed $n$ defined in~\eqref{A1}. As shown explicitly in Eqs.~\eqref{quadH}--\eqref{317} below, the part generated by the infinite spin sum behaves, in the limit $z\to0$ and $\bar z\to1$, as
\begin{equation}
    z^{\Delta_\CO+n} (1-\bar z)^{\Delta_\CO+\rho}\log^2 z .
\end{equation}
Since in this limit we have $u\simeq z$ and $v\simeq1-\bar z$, this gives
\begin{equation}
    \mathcal{G}^{(4)}(u,v)\sim u^{\Delta_\CO+n}v^{\Delta_\CO+\rho}\log^2 u .
\end{equation}
Crossing symmetry then maps this behavior to
\begin{equation}\label{crossedlog2}
    \mathcal{G}^{(4)}(u,v)\sim u^{2\Delta_\CO+\rho}v^n \log^2 v .
\end{equation}
The power of $u$ in \eqref{crossedlog2} cannot come from the direct-channel double-trace sector. Instead, it matches the power expected from a quadruple-trace block \eqref{gquad}.  Thus we see that crossing symmetry forces the exchange of quadruple-trace operators in the $\mathcal{O}\times\mathcal{O}$ OPE. This will become even more clear when we discuss the diagrammatic representation.

\subsubsection{OPE coefficient} \label{subopecoef}

We now want to find the leading OPE coefficient $b_{q,\ell}^{(0)}$ for $q=0$. Our strategy is to apply the Lorentzian inversion formula to $\mathcal{G}^{(4)}(u,v)$ in the crossed channel. We then identify the poles of $c_{\Delta,\ell}$ at the positions of the quadruple-trace operators and read off their residues. For this purpose, the only relevant contribution is the term displayed in~\eqref{dtlog}. Using the functions $H_n^{(m)}(z,\bar z)$ defined in \eqref{A1} the contribution in~\eqref{dtlog} can be written as
\begin{equation}\label{quadH}
    \frac{1}{8}\log^2(z\bar z) \sum_{n,\rho} c_{n,\rho}\,  H_n^{(2\Delta_\CO+\rho)}(z,\bar z).
\end{equation}

We are interested in the behavior of this expression in the $\bar z\to 1$ limit. In the decomposition of $H_n^{(m)}$ in~\eqref{decH}, only the second term is relevant for extracting the crossed-channel singularity that gives rise to the leading quadruple-trace pole. The first term has the behavior of an individual conformal block near $\bar z=1$, and therefore does not generate the behaviour of interest. Thus we focus on
\begin{equation}
    -\frac{1}{8}\frac{\log^2(z\bar z)}{z-\bar z}\sum_{n,\rho} c_{n,\rho}\,z^{\tau_n/2} F_{\frac{\tau_n-2}{2}}(z)\,\tilde{H}_n^{(2\Delta_\CO+\rho)}(\bar z).
\end{equation}
Near $\bar z=1$, we may replace $\tilde{H}_n$ by $P_n$, which is defined in \eqref{ppn}. Since the other terms again have the behavior of individual conformal blocks in this limit. Doing so we find
\begin{equation}\label{317}
    \mathcal{G}^{(4)}(z,\bar z)\supset -\frac{1}{8} \frac{\log^2 z}{z-1} \sum_n c_{n,0}\,z^{\tau_n/2} F_{\frac{\tau_n-2}{2}}(z)\, \frac{Y_n} {\left[(1-\Delta_\CO)_{2\Delta_\CO}\right]^2}(1-\bar z)^{\Delta_\CO}+\mathcal{O}\!\left((1-\bar z)^{\Delta_\CO+1}\right).
\end{equation}
Here the coefficients $Y_n$ are defined in~\eqref{yn}. Note that at leading order in $(1-\bar z)$ expansion, only the $\rho=0$ term contributes. In other words, only the leading term in the large-spin expansion of $\gamma^{(2)}_{n,\ell}$ is needed to extract $b^{(0)}_{0,\ell}$.

Before proceeding, let us comment on the range of validity of this expression. As explained in the Appendix \ref{appxA}, the relation used above holds provided
\begin{equation}
    \Delta_\CO-m \notin \mathbb{Z}_{\leq 0}.
\end{equation}
In the present case $m=2\Delta_\CO+\rho$, and therefore this condition excludes integer values of $\Delta_\CO$. This is precisely the case in which the double-trace and quadruple-trace towers can mix with each other. In what follows then we assume that $\Delta_\CO$ is non-integer. 

We now use crossing symmetry to find the crossed image of~\eqref{317}. It is convenient to define the crossing operator $\mathsf S$ by
\begin{equation}\label{crossingop}
    \mathsf S[f(z,\bar z)] =\left( \frac{z\bar z}{(1-z)(1-\bar z)} \right)^{\Delta_\CO} f(1-\bar z,1-z).
\end{equation}
Under this transformation, the limit $\bar z\to 1$ is mapped to the limit $z\to 0$. Acting with $\mathsf S$ on~\eqref{317}, and keeping only the leading term in $z$ we obtain
\begin{equation} \label{crossg4}
    \mathcal{G}^{(4)}(z,\bar z)\supset\frac{1}{8}\frac{\log^2(1-\bar z)}{\bar z^{1-\Delta_\CO}}\,\frac{z^{2\Delta_\CO}}{\left[(1-\Delta_\CO)_{2\Delta_\CO}\right]^2} \sum_nc_{n,0}\, (1-\bar z)^{n} F_{\frac{\tau_n-2}{2}}(1-\bar z)\, Y_n .
\end{equation}

We may now insert~\eqref{crossg4} into the inversion integral \eqref{LIF}. In the limit that we are taking, the $z$ and $\bar z$ integrals decouple. The $z$-integral takes the form
\begin{equation}
    \int_0^1 dz\, z^{\frac{\ell-\Delta}{2}-1+2\Delta_\CO} =-\frac{2}{\Delta-(4\Delta_\CO+\ell)} .
\end{equation}
Therefore we see that the pole is located at the position of the leading-twist quadruple-trace operator. The residue of this pole is determined by the remaining $\bar z$-integral.

We evaluate the $\bar z$-integral by expanding the summand in~\eqref{crossg4} around $\bar z=1$ and integrating term by term, following the same method as in~\cite{BubblesAdS}. This expansion gives the large-spin expansion of the final result in powers of $1/\ell$. Therefore, to obtain the first few terms in the large-spin expansion of $b^{(0)}_{0,\ell}$, it is sufficient to know the coefficients $c_{n,0}$ for the corresponding finite range of $n$.

The coefficients $c_{n,0}$ can be extracted from \eqref{gam242}. Namely we have
\begin{equation}  \label{cn0}
    c_{n,0}= \zeta_n^2 = 4(1-2\Delta_{\mathcal O})^2 \left[ 2\Delta_{\mathcal O}^2 +(4n-5)\Delta_{\mathcal O} +2n^2-6n+3 \right]^2  \frac{\Gamma(2\Delta_{\mathcal O}+n-3)^2}{\Gamma(n+1)^2}.
\end{equation}

For half-integer values of $\Delta_\CO$, the sum over $n$ in \eqref{crossg4} can be performed exactly. For example, for $\Delta_\CO=\frac{5}{2}$, we find
\begin{equation}
    \sum_n c_{n,0}\, (1-\bar z)^n F_{n+\frac{3}{2}}(1-\bar z)\, Y_n = -\frac{ 36\left[ 3\bar z\bigl(\bar z(43\bar z-365)+725\bigr)-1225 \right]}{\bar z^{9/2}} .
\end{equation}
In this case the remaining $\bar z$-integral in the inversion formula can be performed exactly, without expanding around $\bar z=1$. So the resulting leading quadruple-trace OPE coefficient for $\Delta_\CO=\frac{5}{2}$ is
\begin{equation}
    b^{(0)}_{0,\ell} = \frac{ \sqrt{\pi}\,2^{5-2\ell}(2\ell+9)^2\bigl(\ell(\ell+9)+23\bigr)^2 \Gamma(\ell+1) }{ 225 (\ell+5)(\ell+6)(\ell+7)(\ell+8)\Gamma\!\left(\ell+\frac{9}{2}\right) } .
\end{equation}

For generic non-integer $\Delta_\CO$, the leading large-spin behavior is
\begin{equation}
 b^{(0)}_{0,\ell} =\frac{\sqrt{\pi}\, 2^{-4\Delta_\CO-2\ell+1} \Gamma(2\Delta_\CO)^2}{ \ell^{3/2} \left[(1-\Delta_\CO)_{2\Delta_\CO}\right]^2}\left[ 1+\mathcal{O}\!\left(\frac{1}{\ell}\right) \right].
\end{equation}
Higher-order terms in the $1/\ell$ expansion can be obtained systematically by keeping more terms in the expansion of the summand in~\eqref{crossg4} around $\bar z=1$. Thus the leading large-spin double-trace data are sufficient to fix the leading-twist quadruple-trace OPE coefficient to all orders in the large-spin expansion. For special half-integer values of $\Delta_\CO$, the sum over $n$ can even be performed in closed form, giving an exact expression for $b^{(0)}_{0,\ell}$.

\subsubsection{Anomalous dimension} \label{subanom}

For quadruple-trace operators, the function $c_{\Delta,\ell}$ in the inversion formula \eqref{LIF} has the following expansion
\begin{equation}
    c_{\Delta,\ell} = -\frac{1}{N^8} \frac{b^{(0)}_{q,\ell}}{\Delta-(4\Delta_\CO+2q+\ell)}  -\frac{1}{N^{10}} \left[ \frac{b^{(1)}_{q,\ell}} {\Delta-(4\Delta_\CO+2q+\ell)} +\frac{b^{(0)}_{q,\ell}\delta^{(1)}_{q,\ell}}{\left(\Delta-(4\Delta_\CO+2q+\ell)\right)^2}\right] +\cdots .
\end{equation}
So the anomalous dimension $\delta^{(1)}_{q,\ell}$ should be extracted from the double pole at the quadruple-trace position.

The relevant source of this double pole is the following term in the double-trace sector at order $N^{-10}$:
\begin{equation} \label{g5}
    \mathcal{G}^{(5)}(u,v)  \supset \sum_{n,\ell} \frac{1}{4}\, a^{(0)}_{n,\ell}\,  u^{\Delta_\CO+n}\,  \gamma^{(2)}_{n,\ell}  \gamma^{(3)}_{n,\ell}\big|_{\log J} \left(\log u+\frac{\partial}{\partial n}\right)^2 \tilde{g}_{n,\ell}(u,v).
\end{equation}
Note that the double pole is controlled only by the  part of $\gamma^{(3)}_{n,\ell}$ which is proportional to $\log J$ (See \eqref{wlog}). Therefore, the knowledge of this piece is sufficient to determine the quadruple-trace anomalous dimension $\delta^{(1)}_{q,\ell}$.

As in the analysis of the OPE coefficient, we restrict our attention to the lowest-twist quadruple-trace operators. Therefore, in the large-spin expansion
\begin{equation}
    \gamma^{(2)}_{n,\ell}\gamma^{(3)}_{n,\ell}\Big|_{\log J} = \sum_{\rho\geq 0}  \frac{w_{n,\rho}}{J^{4\Delta_\CO+2\rho}} ,
\end{equation}
only the term with $\rho=0$ is needed. From \eqref{gam242} and \eqref{gam3lll}, we then have
\begin{equation}
    w_{n,0}=-\zeta_n^2=-c_{n,0},
\end{equation}
where $c_{n,0}$ is given in \eqref{cn0}.

Using the discussion in Appendix~\ref{appxA}, in particular \eqref{wlog}, the relevant part of \eqref{g5} in the limit $\bar z\to 1$ becomes
\begin{equation}
    \mathcal{G}^{(5)}(u,v) \supset \frac{1}{8\left[(1-\Delta_\CO)_{2\Delta_\CO}\right]^2} \frac{\log^2(z\bar z)}{z-1} \sum_n w_{n,0}Y_n\, z^{\tau_n/2} F_{\frac{\tau_n-2}{2}}(z)\, (1-\bar z)^{\Delta_\CO} \log(1-\bar z).
\end{equation}

We now use crossing symmetry in the form prescribed in \eqref{crossingop}. In the crossed channel, and in the limit $z\to 0$, this gives
\begin{equation}\label{anom1}
    \mathcal{G}^{(5)}(u,v)  \supset \frac{\log^2(1-\bar z)}{8\left[(1-\Delta_\CO)_{2\Delta_\CO}\right]^2} \frac{z^{2\Delta_\CO}\log z}{\bar z^{1-\Delta_\CO}} \sum_nc_{n,0}Y_n\,  (1-\bar z)^n F_{\frac{\tau_n-2}{2}}(1-\bar z).
\end{equation}
In this limit, the $z$ and $\bar z$ integrals in the inversion formula decouple. The $z$-integral gives
\begin{equation}
    \int_0^1 dz\,z^{\frac{\ell-\Delta}{2}-1+2\Delta_\CO} \log z = -\frac{4}  {\left(\Delta-(4\Delta_\CO+\ell)\right)^2}.
\end{equation}
This produces the expected double pole at the position of the lowest-twist quadruple-trace operator. Now it remains to evaluate the $\bar z$-integral, but comparing \eqref{anom1} with \eqref{crossg4} we see that the $\bar z$ integral is identical to the one for the OPE coefficient. Therefore we find
\begin{equation}
    \delta_{0,\ell}^{(1)}=2.
\end{equation}

\subsection{$\phi^3$ theory} \label{subphi3}

We now turn to a bulk theory with a cubic interaction. In the CFT this means that $\CO$ itself appears in the $\CO\times\CO$ OPE, so the $\mathbb Z_2$ symmetry of the previous subsection is absent. Schematically,
\begin{equation}\label{333}
\CO \times \CO = \mathds{1} + [\CO\CO]_{n,\ell} + \frac{1}{N}\CO + \sum_{k=3}^{\sigma} \frac{1}{N^k}[\CO^k]_{r,\ell,I} +\cdots .
\end{equation}
Here $[\CO^k]_{r,\ell,I}$ denotes a $k$-trace family built out of $\CO$, $r$ labels the twist level, and $I$ is a possible degeneracy label within that family. Moreover, we assume that, up to order $1/N^\sigma$ in the OPE, different trace families do not mix with each other. 

The scaling dimension and squared OPE coefficient of a $k-$trace operator can be expanded in large $N$ as
\begin{equation}
\Delta_{[\CO^k]_{r,\ell,I}} = k\Delta_\CO+2r+\ell + \frac{\eta^{(1)}_{k,r,\ell,I}}{N^2} +\cdots ,
\end{equation}
\begin{equation}
a_{[\CO^k]_{r,\ell,I}} = \frac{\mathfrak a^{(0)}_{k,r,\ell,I}}{N^{2k}} +\cdots .
\end{equation}
As before, we can sum over the degeneracy label and define
\begin{equation}
\mathfrak a^{(0)}_{k,r,\ell} \equiv \sum_I \mathfrak a^{(0)}_{k,r,\ell,I}.
\end{equation}

Our goal is to find this coefficient for the leading twist $r=0$. The strategy is the same as in the $\phi^4$ analysis. We start from the double-trace part of the conformal block expansion and isolate the terms which, after the infinite spin sum and crossing, have the correct small-$z$ behavior to be interpreted as the exchange of $k$-trace operators.

At order $N^{-2k}$ in the correlator, several double-trace terms have this property. Among them, the term with the highest power of $\log u$ is
\begin{equation}\label{general}
\mathcal G^{(k)}(u,v) \supset \frac{1}{2^k k!} \sum_{n,\ell} a^{(0)}_{n,\ell} u^{\Delta_\CO+n} \left(\gamma^{(1),\mathrm{as}}_{n,\ell}\right)^k \log^k u\, \tilde g_{n,\ell}(u,v).
\end{equation}
Here $\gamma^{(1),\mathrm{as}}_{n,\ell}$ denotes the asymptotic part of the tree-level anomalous dimension, given in~\eqref{g225} for $n=0$. 

We expand
\begin{equation}
\left(\gamma^{(1),\mathrm{as}}_{n,\ell}\right)^k = \sum_{\rho\geq0} \frac{d_{n,\rho}}{J^{k\Delta_\CO+2\rho}}.
\end{equation}
Using the functions $H_n^{(m)}(z,\bar z)$ defined in
Appendix~\ref{appxA}, the contribution in~\eqref{general} can be written as
\begin{equation}
    \mathcal G^{(k)}(z,\bar z)  \supset  \frac{\log^k(z\bar z)}{2^k k!} \sum_{n,\rho} d_{n,\rho}\, H_n^{\left(\frac{k\Delta_\CO}{2}+\rho\right)}(z,\bar z).
\end{equation}
Near $\bar z=1$, the relevant part of the spin sum is the second term in~\eqref{decH}, with $\tilde H_n^{(m)}(\bar z)$ replaced by $P_n^{(m)}(\bar z)$. For generic values of $\Delta_\CO$ for which \eqref{ppn} applies, the leading behavior as $\bar z\to1$ selects the term with $\rho=0$, since the terms with $\rho>0$ are suppressed by additional powers of $1-\bar z$. We then obtain around $\bar z=1$:
\begin{equation}
    \mathcal G^{(k)}(z,\bar z) \supset -\frac{\log^k z}{2^k k!(z-1)}\frac{(1-\bar z)^{\left(\frac{k}{2}-1\right)\Delta_\CO}}{\left[(1-\Delta_\CO)_{\frac{k\Delta_\CO}{2}}\right]^2}\sum_n d_{n,0}Y_n\,z^{\tau_n/2}F_{\frac{\tau_n-2}{2}}(z)+\cdots .
\end{equation}
Applying crossing symmetry and keeping the leading term as $z\to0$ gives
\begin{equation}\label{new1}
    \mathcal G^{(k)}(z,\bar z) \supset \frac{z^{\frac{k\Delta_\CO}{2}}\log^k(1-\bar z)}{2^k k!\,\bar z^{1-\Delta_\CO}\left[(1-\Delta_\CO)_{\frac{k\Delta_\CO}{2}}\right]^2}\sum_n d_{n,0}Y_n\,(1-\bar z)^nF_{\frac{\tau_n-2}{2}}(1-\bar z) .
\end{equation}
Now performing the $z$-integral of the inversion formula gives: 
\begin{equation}
    \int_0^1 dz\, z^{\frac{\ell-\Delta}{2}-1+\frac{k\Delta_\CO}{2}}= -\frac{2}{\Delta-(k\Delta_\CO+\ell)}.
\end{equation}

More generally, in the crossed channel, the term labelled by $\rho$ would be proportional to $z^{k\Delta_\CO/2+\rho}$ and would therefore produce a pole at $\Delta=k\Delta_\CO+2\rho+\ell$. Thus, $\rho=0$ selects the leading-twist $k$-trace family.

To extract the leading large-spin behavior of the residue, we expand the remaining summand around $\bar z=1$. At leading order this picks the $n=0$ term in~\eqref{new1}. Moreover, before inserting into the inversion formula, we note that
\begin{equation}
    \dDisc\!\left[\log^k(1-\bar z)\right] =2\pi^2 k(k-1)\log^{k-2}(1-\bar z)+\cdots .
\end{equation}
As was shown in~\cite{BubblesAdS}, powers of $\log(1-\bar z)$ in the inversion integral generate the corresponding powers of $\log\ell$ in the large-spin expansion. Therefore, although~\eqref{general} is not the only contribution to the $k$-trace OPE coefficient, it uniquely determines its leading $\log^{k-2}\ell$ behavior. Performing the remaining $\bar z$-integral, we find

\begin{equation}\label{ak0l}
\mathfrak a^{(0)}_{k,0,\ell\gg1} =\frac{(-1)^k\, 2^{1-k\Delta_\CO-2\ell}\sqrt{\pi}\,(\tilde\gamma_0)^k}{(k-2)!\,\left[(1-\Delta_\CO)_{\frac{k\Delta_\CO}{2}}\right]^2}\,\frac{\log^{k-2}\ell}{\ell^{3/2}}+\cdots .
\end{equation}

where $\tilde\gamma_0$ can be read off from \eqref{g225} and is given by
\begin{equation}
  \tilde\gamma_0=  -\frac{2a_{\Delta_\mathcal{O},0}\Gamma(\Delta_\mathcal{O})^3}{\Gamma\!\left(\frac{\Delta_\mathcal{O}}{2}\right)^4}.
\end{equation}
We remind that this result is written under the no-mixing assumption stated earlier. Namely for $k\leq\sigma$.

\section{Diagrammatic representation}\label{sec4}

In this section we introduce a diagrammatic notation for organizing the terms that appear in the conformal block expansion. These diagrams keep track of the combinations of OPE data multiplying a given conformal block.

The motivation comes from the analysis of bubble diagrams \cite{BubblesAdS}. In that case, it was observed that a part of the conformal block expansion that is completely fixed by tree-level data is related to the AdS Bubble diagrams with maximal unitarity cuts. By ``maximal''  we mean that the bubble diagram is cut iteratively through all its internal propagators. Here we extend this observation into a more systematic dictionary. Namely, terms in the conformal block expansion are represented by cut diagrams. The cuts need not be maximal. We use vertical cuts to denote direct-channel expansion and horizontal cuts to denote the crossed-channel expansion.

This construction is closely related to existing holographic unitarity methods~\cite{Meltzer:2019nbs}. In these methods, loop-level CFT data are reconstructed from products of lower-order OPE data. The bulk formulation of~\cite{Meltzer:2019nbs} gives this factorization a direct diagrammatic meaning. Namely, a cut puts internal bulk propagators on shell and factorizes a Witten diagram into lower-loop subdiagrams.

Our construction starts instead directly from the conformal block expansion. When an internal cut separates a diagram into subdiagrams, the associated OPE data factorize into lower-loop data according to a product rule that we will introduce. In this sense, our internal cuts, whether drawn vertically or horizontally, have the same algebraic behavior as unitarity cuts. Since the two orientations correspond to different OPE channels, this suggests that internal vertical and horizontal cuts may be related to bulk unitarity cuts in the corresponding channels. However, in this work, we do not define the cuts by an on-shell operation in the bulk, nor do we establish that the internal cuts introduced here are AdS unitarity cuts.

Regardless of this possible bulk interpretation, this diagrammatic notation has two useful purposes on the CFT side. First, it provides a more detailed bookkeeping of the CFT data contributing at a given order in $1/N$. Second, it makes it possible to identify crossing-closed classes of diagrams, namely subsets of terms that are mapped into themselves under crossing.

Let us recall the conformal-block expansion of the four-point function,
\begin{equation} \label{diag11}
    \mathcal G(u,v) =\sum_{\Delta,\ell}a_{\Delta,\ell}\,u^{\frac{\Delta-\ell}{2}} \tilde g_{\Delta,\ell}(u,v) .
\end{equation}
In the large-$N$ expansion we write
\begin{equation} \label{diag12}
    \mathcal G(u,v) = \mathcal G^{(0)}(u,v)+{1\over N^2}\mathcal G^{(1)}(u,v)+{1\over N^4}\mathcal G^{(2)}(u,v)+\cdots .
\end{equation}
This is the CFT counterpart of the bulk loop expansion. At each order in $1/N$, our goal is to associate the different terms in the conformal-block expansion with diagrams carrying specified cuts.

\subsection{Disconnected contribution}
At leading order the correlator is the generalized-free-field correlator,
\begin{equation}\label{discexp1}
    \mathcal G^{(0)}(u,v)  =  1+u^{\Delta_\CO}  +\left(\frac{u}{v}\right)^{\Delta_\CO}=1+\sum_{n,\ell}a^{(0)}_{n,\ell} u^{\Delta_\CO+n}\tilde g_{n,\ell}(u,v),
\end{equation}
where the sum runs over the double-trace operators $[\CO\CO]_{n,\ell}$. In this decomposition, the first term is the identity contribution, and the remaining disconnected pieces are reproduced by the exchange of the double-trace tower in the direct channel.

In terms of Witten diagrams, we can represent the three disconnected contributions by
\begin{equation}
\mathcal{G}^{(0)}(u,v)
=
\begin{tikzpicture}[baseline=-0.5ex, scale=0.9]
    \draw[thick] (0,0) circle (0.55);
    \draw[thick] (-0.25,-0.50) -- (-0.25,0.50);
    \draw[thick] (0.12,-0.53) -- (0.12,0.53);
\end{tikzpicture}
\;+\;
\begin{tikzpicture}[baseline=-0.5ex, scale=0.9]
    \draw[thick] (0,0) circle (0.55);
    \draw[thick] (-0.55,0.12) -- (0.55,0.12);
    \draw[thick] (-0.55,-0.20) -- (0.55,-0.20);
\end{tikzpicture}
\;+\;
\begin{tikzpicture}[baseline=-0.5ex, scale=0.9]
    \draw[thick] (0,0) circle (0.55);

    \draw[thick] (-0.55,0.12) -- (0.12,0.12);
    \draw[thick] (0.12,0.12) -- (0.53,-0.18);

    \draw[thick] (-0.55,-0.20) -- (0.15,-0.20);
    \draw[thick] (0.15,-0.20) -- (0.29,-0.09);
    \draw[thick] (0.39,-0.01) -- (0.53,0.12);
\end{tikzpicture}
\, .
\end{equation}
Note that the first diagram admits a vertical cut which does not intersect any propagator. We interpret this cut as the exchange of the identity operator in the direct-channel OPE, and we make the following diagrammatic identification
\begin{equation}
\left(
\,\begin{tikzpicture}[
    baseline=-0.5ex,
    scale=0.85,
    line cap=round,
    line join=round
]
    \draw[thick] (-0.35,-0.80) -- (-0.35,0.80);
    \draw[thick] ( 0.35,-0.80) -- ( 0.35,0.80);

    \foreach \y in {-0.70,-0.42,-0.14,0.14,0.42,0.70}{
        \fill (0,\y) circle (0.040);
    }
\end{tikzpicture}\,
\right)
=1\, .
\label{diag13}
\end{equation}
The other two diagrams admit a two-particle cut. Such a cut represents the exchange of two-particle states, dual to the double-trace operators $[\CO\CO]_{n,\ell}$. Thus, for a fixed exchanged double-trace operator, we make the identification
\begin{equation}\label{diag1}
\left[
\left(
\,\begin{tikzpicture}[
    baseline=-0.6ex,
    scale=0.85,
    line cap=round,
    line join=round
]
    \draw[thick] (-0.85,0.25) -- (0.85,0.25);
    \draw[thick] (-0.85,-0.25) -- (0.85,-0.25);

    \foreach \y in {-0.65,-0.39,-0.13,0.13,0.39,0.65}{
        \fill (0,\y) circle (0.040);
    }
\end{tikzpicture}\,
\right)
+
\left(
\,\begin{tikzpicture}[
    baseline=-0.6ex,
    scale=0.85,
    line cap=round,
    line join=round
]

    \draw[thick] (-0.85, 0.25) -- (-0.18, 0.25);
    \draw[thick] (-0.85,-0.25) -- (-0.18,-0.25);

    \draw[thick] (-0.18,-0.25) -- (0.23,-0.05);
    \draw[thick] (0.43,0.05) -- (0.85,0.25);

    \draw[thick] (-0.18,0.25) -- (0.85,-0.25);

    \foreach \y in {-0.65,-0.39,-0.13,0.13,0.39,0.65}{
        \fill (-0.40,\y) circle (0.040);
    }
\end{tikzpicture}\,
\right)
\right]_{n,\ell}
=
a^{(0)}_{n,\ell}\, .
\end{equation}

This motivates the following shorthand. A dot-cut keeps track of a fixed pairing of the two particles across the cut. For identical external operators, we want the cut to include the two possible pairings of the external legs on one side of the cut. So we define the dash-cut in the following way,
\begin{equation}
\left(
\,\begin{tikzpicture}[
    baseline=-0.5ex,
    scale=0.9,
    line cap=round,
    line join=round
]

\draw[thick,pattern=north east lines] (0,0) circle (0.50);

    \draw[thick] (-1.10, 0.60) -- (-0.42, 0.27);
    \draw[thick] (-1.10,-0.60) -- (-0.42,-0.27);

    \draw[thick] (0.42, 0.27) -- (1.10, 0.60);
    \draw[thick] (0.42,-0.27) -- (1.10,-0.60);

    \draw[thick,dash pattern=on 5pt off 3pt] (0,-1.10) -- (0,1.10);
\end{tikzpicture}\,
\right)
=
\frac{1}{\mathfrak{s}}
\Bigg[
\left(
\,\begin{tikzpicture}[
    baseline=-0.5ex,
    scale=0.9,
    line cap=round,
    line join=round
]
    \draw[thick,pattern=north east lines] (0,0) circle (0.50);

    \draw[thick] (-1.10, 0.60) -- (-0.42, 0.27);
    \draw[thick] (-1.10,-0.60) -- (-0.42,-0.27);

    \draw[thick] (0.42, 0.27) -- (1.10, 0.60);
    \draw[thick] (0.42,-0.27) -- (1.10,-0.60);

    \foreach \y in {-0.92,-0.58,-0.24,0.10,0.44,0.78}{
        \fill (0,\y) circle (0.040);
    }
\end{tikzpicture}\,
\right)
+
\left(
\,\begin{tikzpicture}[
    baseline=-0.5ex,
    scale=0.9,
    line cap=round,
    line join=round
]

\draw[thick,pattern=north east lines] (0,0) circle (0.50);

    \draw[thick] (-1.10, 0.60) -- (-0.42, 0.27);
    \draw[thick] (-1.10,-0.60) -- (-0.42,-0.27);

    \draw[thick] (0.42, 0.27) -- (0.70, 0.27);
    \draw[thick] (0.42,-0.27) -- (0.70,-0.27);

    \draw[thick] (0.70,-0.27) -- (0.78,-0.09);
    \draw[thick] (0.88, 0.13) -- (1.10,0.60);

    \draw[thick] (0.70,0.27) -- (1.10,-0.60);

    \foreach \y in {-0.92,-0.58,-0.24,0.10,0.44,0.78}{
        \fill (0,\y) circle (0.040);
    }
\end{tikzpicture}\,
\right)
\Bigg] .
\label{diag14}
\end{equation}
where
\begin{equation}\label{diag15}
\mathfrak{s}=
\begin{cases}
2, & \text{if the two dot-cut diagrams on the right-hand side coincide},\\1, & \text{otherwise}.
\end{cases}
\end{equation}
So the factor $\mathfrak{s}$ removes the double counting when the two outgoing pairings give the same diagram.

Using the dash-cut definition~\eqref{diag14}, the relation \eqref{diag1} can be written more simply as
\begin{equation}
\left(
\,\begin{tikzpicture}[
    baseline=-0.5ex,
    scale=0.85,
    line cap=round,
    line join=round
]
    \draw[thick] (-0.85, 0.25) -- (0.85, 0.25);
    \draw[thick] (-0.85,-0.25) -- (0.85,-0.25);

    \draw[thick,dash pattern=on 5pt off 3pt] (0,-0.70) -- (0,0.70);
\end{tikzpicture}\,
\right)_{n,\ell}= a^{(0)}_{n,\ell}\, .
\label{diag16}
\end{equation}
From now on, unless stated otherwise, by a cut we mean a dash-cut. When we need to keep track of a fixed pairing across the cut, we will explicitly refer to it as a dot-cut.

\paragraph{Two-fold notation.} We use two related diagrammatic notations. A diagram inside parentheses denotes the OPE data factor multiplying a conformal block. The same diagram enclosed by a circle denotes the corresponding contribution to the correlator, after summing over $n$ and $\ell$. Thus we write
\begin{equation}
\begin{tikzpicture}[
    baseline=-0.5ex,
    scale=0.75,
    line cap=round,
    line join=round
]
  \draw[thick] (0,0) circle (1.15);

  \draw[thick] (-1.09,0.28)--(1.09,0.28);
  \draw[thick] (-1.09,-0.28)--(1.09,-0.28);

  \draw[thick,dash pattern=on 5pt off 3pt] (0,-0.95)--(0,0.95);
\end{tikzpicture}
=
\sum_{n,\ell}
u^{\Delta_\CO+n}
\left(
\,\begin{tikzpicture}[
    baseline=-0.5ex,
    scale=0.55,
    line cap=round,
    line join=round
]

  \draw[thick] (-0.85,0.25)--(0.85,0.25);
  \draw[thick] (-0.85,-0.25)--(0.85,-0.25);

  \draw[thick,dash pattern=on 5pt off 3pt] (0,-0.70)--(0,0.70);
\end{tikzpicture}\,\right)_{n,\ell}\tilde g_{n,\ell}(u,v)\, .
\label{diag17}
\end{equation}
For the identity contribution, the corresponding OPE coefficient is normalized to one, as in \eqref{diag13}. Therefore the disconnected correlator can be written diagrammatically as
\begin{equation}
\mathcal{G}^{(0)}(u,v)
=
\begin{tikzpicture}[
    baseline=-0.5ex,
    scale=0.75,
    line cap=round,
    line join=round
]
  \draw[thick] (0,0) circle (1.15);

  \draw[thick] (-0.32,-1.10)--(-0.32,1.10);
  \draw[thick] (0.32,-1.10)--(0.32,1.10);

  \draw[thick,dash pattern=on 5pt off 3pt] (0,-0.85)--(0,0.85);
\end{tikzpicture}
\;+\;
\begin{tikzpicture}[
    baseline=-0.5ex,
    scale=0.75,
    line cap=round,
    line join=round
]
  \draw[thick] (0,0) circle (1.15);

  \draw[thick] (-1.10,0.28)--(1.10,0.28);
  \draw[thick] (-1.10,-0.28)--(1.10,-0.28);

  \draw[thick,dash pattern=on 5pt off 3pt] (0,-0.95)--(0,0.95);
\end{tikzpicture}
\, .
\label{diag18}
\end{equation}
The first term represents the identity exchange, and the second term represents the contribution of the double-trace tower in the direct channel.

\paragraph{Crossing.}
The crossing operator $\mathsf S$ defined in \eqref{crossingop} acts on the circled diagrams, as these diagrams denote contributions to the correlator. Diagrammatically, we represent this action by a rotation by $\pi/2$, which also exchanges vertical and horizontal cuts.

At the level of OPE data, such rotation does not change the identification. For instance,
\begin{equation}
\left(
\,\begin{tikzpicture}[
    baseline=-0.5ex,
    scale=0.85,
    line cap=round,
    line join=round
]

  \draw[thick] (-0.85, 0.25)--(0.85, 0.25);
  \draw[thick] (-0.85,-0.25)--(0.85,-0.25);

  \draw[thick,dash pattern=on 5pt off 3pt] (0,-0.70)--(0,0.70);
\end{tikzpicture}\,
\right)_{n,\ell}
=
\left(
\,\begin{tikzpicture}[
    baseline=-0.5ex,
    scale=0.85,
    line cap=round,
    line join=round
]

  \draw[thick] (-0.25,-0.85)--(-0.25,0.85);
  \draw[thick] ( 0.25,-0.85)--( 0.25,0.85);

  \draw[thick,dash pattern=on 5pt off 3pt] (-0.70,0)--(0.70,0);
\end{tikzpicture}\,
\right)_{n,\ell}
=
a^{(0)}_{n,\ell}\, .
\label{diag19}
\end{equation}
But for circled diagrams, we have for instance,
\begin{equation}
\begin{tikzpicture}[
    baseline=-0.5ex,
    scale=0.72,
    line cap=round,
    line join=round
]
  \draw[thick] (0,0) circle (1.10);

  \draw[thick] (-0.30,-1.05)--(-0.30,1.05);
  \draw[thick] ( 0.30,-1.05)--( 0.30,1.05);

  \draw[thick,dash pattern=on 5pt off 3pt] (-0.85,0)--(0.85,0);
\end{tikzpicture}
=
\mathsf S
\left[
\begin{tikzpicture}[
    baseline=-0.5ex,
    scale=0.72,
    line cap=round,
    line join=round
]
  \draw[thick] (0,0) circle (1.10);

  \draw[thick] (-1.05, 0.28)--(1.05, 0.28);
  \draw[thick] (-1.05,-0.28)--(1.05,-0.28);

  \draw[thick,dash pattern=on 5pt off 3pt] (0,-0.85)--(0,0.85);
\end{tikzpicture}
\right]\, .
\label{diag20}
\end{equation}

With this notation, crossing symmetry equation for the disconnected correlator can be written diagrammatically as
\begin{equation}
\begin{tikzpicture}[
    baseline=-0.5ex,
    scale=0.72,
    line cap=round,
    line join=round
]
  \draw[thick] (0,0) circle (1.10);

  \draw[thick] (-0.30,-1.05)--(-0.30,1.05);
  \draw[thick] ( 0.30,-1.05)--( 0.30,1.05);

  \draw[thick,dash pattern=on 5pt off 3pt] (0,-0.85)--(0,0.85);
\end{tikzpicture}
\;+\;
\begin{tikzpicture}[
    baseline=-0.5ex,
    scale=0.72,
    line cap=round,
    line join=round
]
  \draw[thick] (0,0) circle (1.10);

  \draw[thick] (-1.05, 0.28)--(1.05, 0.28);
  \draw[thick] (-1.05,-0.28)--(1.05,-0.28);

  \draw[thick,dash pattern=on 5pt off 3pt] (0,-0.85)--(0,0.85);
\end{tikzpicture}
=
\begin{tikzpicture}[
    baseline=-0.5ex,
    scale=0.72,
    line cap=round,
    line join=round
]
  \draw[thick] (0,0) circle (1.10);

  \draw[thick] (-0.30,-1.05)--(-0.30,1.05);
  \draw[thick] ( 0.30,-1.05)--( 0.30,1.05);

  \draw[thick,dash pattern=on 5pt off 3pt] (-0.85,0)--(0.85,0);
\end{tikzpicture}
\;+\;
\begin{tikzpicture}[
    baseline=-0.5ex,
    scale=0.72,
    line cap=round,
    line join=round
]
  \draw[thick] (0,0) circle (1.10);

  \draw[thick] (-1.05, 0.28)--(1.05, 0.28);
  \draw[thick] (-1.05,-0.28)--(1.05,-0.28);

  \draw[thick,dash pattern=on 5pt off 3pt] (-0.85,0)--(0.85,0);
\end{tikzpicture}
\, .
\label{diag21}
\end{equation}

\subsection{Higher orders}
Let us now move beyond the disconnected contribution. We want to formulate more general rules for cutting and merging two-particle states, corresponding to double-trace operators in the CFT.

Before doing so, we recall that multi-particle states in AdS are dual to multi-trace operators in the boundary CFT~\cite{Balasubramanian:2001nh,Witten:2001ua}. A diagram may admit cuts through one, two, three, four, or more internal propagators. These cuts therefore signal the exchange of single-trace, double-trace, triple-trace, quadruple-trace, or higher-trace operators in the OPE. We call a cut which intersects $p$-particle propagators a $p$-particle cut.

In this work we focus mainly on general rules for decomposing two-particle cuts. Other types of cuts may also appear, and in those cases we treat the corresponding cut diagrams as building blocks, together with their associated OPE data. 

For example, a tree-level single-trace exchange in $\phi^3$ theory contains a one-particle cut which can be identified as
\begin{equation}
\left(
\,\begin{tikzpicture}[
    baseline=-0.5ex,
    scale=0.9,
    line cap=round,
    line join=round
]
    \fill (-0.45,0) circle (0.04);
    \fill ( 0.45,0) circle (0.04);

    \draw[thick] (-1.10, 0.60) -- (-0.45,0);
    \draw[thick] (-1.10,-0.60) -- (-0.45,0);

  \draw[thick] (-0.45,0) -- (0.45,0);

    \draw[thick] (0.45,0) -- (1.10, 0.60);
    \draw[thick] (0.45,0) -- (1.10,-0.60);

    \draw[thick,dash pattern=on 5pt off 3pt] (0,-0.75) -- (0,0.75);
\end{tikzpicture}\,
\right)
=
a_{\Delta_{\CO},0}\, .
\label{diag22}
\end{equation}

Now let us return to the two-particle cuts. We first consider the case where a two-particle cut goes through the external legs. In that case we make the following identifications:
\begin{equation}
\left(
\,\begin{tikzpicture}[
    baseline=-0.5ex,
    scale=0.85,
    line cap=round,
    line join=round
]
  \draw[thick] (-1.10, 0.60) -- (-0.42, 0.27);
  \draw[thick] (-1.10,-0.60) -- (-0.42,-0.27);
  \draw[thick] ( 0.42, 0.27) -- ( 1.10, 0.60);
  \draw[thick] ( 0.42,-0.27) -- ( 1.10,-0.60);
  \draw[thick,pattern=north east lines] (0,0) circle (0.50);

  \draw[thick,dash pattern=on 5pt off 3pt] (-0.78,-0.75)--(-0.78,0.75);
\end{tikzpicture}\,
\right)^{(0)}_{n,\ell}
\equiv
a^{(0)}_{n,\ell}\,\gamma^{(k)(i)}_{n,\ell},
\qquad
\left(
\,\begin{tikzpicture}[
    baseline=-0.5ex,
    scale=0.85,
    line cap=round,
    line join=round
]
  \draw[thick] (-1.10, 0.60) -- (-0.42, 0.27);
  \draw[thick] (-1.10,-0.60) -- (-0.42,-0.27);
  \draw[thick] ( 0.42, 0.27) -- ( 1.10, 0.60);
  \draw[thick] ( 0.42,-0.27) -- ( 1.10,-0.60);
  \draw[thick,pattern=north east lines] (0,0) circle (0.50);

  \draw[thick,dash pattern=on 5pt off 3pt] (-0.78,-0.75)--(-0.78,0.75);
\end{tikzpicture}\,
\right)^{(1)}_{n,\ell}
\equiv
a^{(k)(i)}_{n,\ell} .
\label{diag23}
\end{equation}
The superscripts $(0)$ and $(1)$ in the diagrams distinguish two different identifications. The label $(i)$ distinguishes inequivalent diagrams contributing to the same OPE data, so we have
\begin{equation}
    a^{(k)}_{n,\ell}=\sum_i a^{(k)(i)}_{n,\ell},
    \qquad
    \gamma^{(k)}_{n,\ell}=\sum_i\gamma^{(k)(i)}_{n,\ell} .
    \label{diag24}
\end{equation}
At tree level, this notation gives the following external two-particle cut identifications:
\begin{equation}
\left(
\,\begin{tikzpicture}[
    baseline=-0.5ex,
    scale=0.8,
    line cap=round,
    line join=round
]
  \draw[thick] (-1.00, 0.55) -- (1.00,-0.55);
  \draw[thick] (-1.00,-0.55) -- (1.00, 0.55);

  \draw[thick,dash pattern=on 5pt off 3pt] (-0.58,-0.72)--(-0.58,0.72);
\end{tikzpicture}\,
\right)^{(0)}_{n,\ell}
\equiv
a^{(0)}_{n,\ell}\,\gamma^{(1)(1)}_{n,\ell},
\qquad
\left(
\,\begin{tikzpicture}[
    baseline=-0.5ex,
    scale=0.8,
    line cap=round,
    line join=round
]
  \draw[thick] (-1.00, 0.55) -- (1.00,-0.55);
  \draw[thick] (-1.00,-0.55) -- (1.00, 0.55);

  \draw[thick,dash pattern=on 5pt off 3pt] (-0.58,-0.72)--(-0.58,0.72);
\end{tikzpicture}\,
\right)^{(1)}_{n,\ell}
\equiv
a^{(1)(1)}_{n,\ell} ,
\label{diag25}
\end{equation}
\begin{equation}
\left(
\,\begin{tikzpicture}[
    baseline=-0.5ex,
    scale=0.8,
    line cap=round,
    line join=round
]
  \draw[thick] (0,0.32)--(0,-0.32);

  \draw[thick] (-1.00, 0.62)--(0, 0.32);
  \draw[thick] ( 0, 0.32)--(1.00, 0.62);
  \draw[thick] (-1.00,-0.62)--(0,-0.32);
  \draw[thick] ( 0,-0.32)--(1.00,-0.62);

  \draw[thick,dash pattern=on 5pt off 3pt] (-0.62,-0.74)--(-0.62,0.74);
\end{tikzpicture}\,
\right)^{(0)}_{n,\ell}
\equiv
a^{(0)}_{n,\ell}\,\gamma^{(1)(2)}_{n,\ell},
\qquad
\left(
\,\begin{tikzpicture}[
    baseline=-0.5ex,
    scale=0.8,
    line cap=round,
    line join=round
]
\draw[thick] (0,0.32)--(0,-0.32);

  \draw[thick] (-1.00, 0.62)--(0, 0.32);
  \draw[thick] ( 0, 0.32)--(1.00, 0.62);
  \draw[thick] (-1.00,-0.62)--(0,-0.32);
  \draw[thick] ( 0,-0.32)--(1.00,-0.62);

  \draw[thick,dash pattern=on 5pt off 3pt] (-0.62,-0.74)--(-0.62,0.74);
\end{tikzpicture}\,
\right)^{(1)}_{n,\ell}
\equiv
a^{(1)(2)}_{n,\ell} ,
\label{diag26}
\end{equation}

\begin{equation}
\left(
\,\begin{tikzpicture}[
    baseline=-0.5ex,
    scale=0.8,
    line cap=round,
    line join=round
]
  \coordinate (L) at (-0.35,0);
  \coordinate (R) at ( 0.35,0);

  \draw[thick] (L)--(R);

  \draw[thick] (-1.00, 0.58)--(L);
  \draw[thick] (-1.00,-0.58)--(L);
  \draw[thick] (R)--(1.00, 0.58);
  \draw[thick] (R)--(1.00,-0.58);

  \draw[thick,dash pattern=on 5pt off 3pt] (-0.65,-0.72)--(-0.65,0.72);
\end{tikzpicture}\,
\right)^{(0)}_{n,\ell}
\equiv
a^{(0)}_{n,\ell}\,\gamma^{(1)(3)}_{n,\ell},
\qquad
\left(
\,\begin{tikzpicture}[
    baseline=-0.5ex,
    scale=0.8,
    line cap=round,
    line join=round
]
  \coordinate (L) at (-0.35,0);
  \coordinate (R) at ( 0.35,0);

\draw[thick] (L)--(R);

  \draw[thick] (-1.00, 0.58)--(L);
  \draw[thick] (-1.00,-0.58)--(L);
  \draw[thick] (R)--(1.00, 0.58);
  \draw[thick] (R)--(1.00,-0.58);

  \draw[thick,dash pattern=on 5pt off 3pt] (-0.65,-0.72)--(-0.65,0.72);
\end{tikzpicture}\,
\right)^{(1)}_{n,\ell}
\equiv
a^{(1)(3)}_{n,\ell} .
\label{diag27}
\end{equation}
The index $i=1,2,3$ labels the three inequivalent tree-level diagrams shown above.\footnote{The separation of the tree-level anomalous dimension into different bulk channels was also considered in~\cite{Meltzer:2019nbs}.}

At one loop, the same rule gives contributions proportional to $a^{(0)}_{n,\ell}\gamma^{(2)}_{n,\ell}$ and $a^{(2)}_{n,\ell}$. For example,
\begin{equation}
\left(
\,\begin{tikzpicture}[
    baseline=-0.5ex,
    scale=0.8,
    line cap=round,
    line join=round
]
  \coordinate (T) at (0,0.28);
  \coordinate (B) at (0,-0.28);

  \draw[thick] (T) .. controls (-0.30,0.08) and (-0.30,-0.08) .. (B);
  \draw[thick] (T) .. controls ( 0.30,0.08) and ( 0.30,-0.08) .. (B);

  \draw[thick] (-1.00, 0.62)--(T);
  \draw[thick] ( 1.00, 0.62)--(T);
  \draw[thick] (-1.00,-0.62)--(B);
  \draw[thick] ( 1.00,-0.62)--(B);

  \draw[thick,dash pattern=on 5pt off 3pt] (-0.62,-0.74)--(-0.62,0.74);
\end{tikzpicture}\,
\right)^{(0)}_{n,\ell}
\equiv
a^{(0)}_{n,\ell}\,\gamma^{(2)(1)}_{n,\ell},
\qquad
\left(
\,\begin{tikzpicture}[
    baseline=-0.5ex,
    scale=0.8,
    line cap=round,
    line join=round
]
  \coordinate (T) at (0,0.28);
  \coordinate (B) at (0,-0.28);

  \draw[thick] (T) .. controls (-0.30,0.08) and (-0.30,-0.08) .. (B);
  \draw[thick] (T) .. controls ( 0.30,0.08) and ( 0.30,-0.08) .. (B);

  \draw[thick] (-1.00, 0.62)--(T);
  \draw[thick] ( 1.00, 0.62)--(T);
  \draw[thick] (-1.00,-0.62)--(B);
  \draw[thick] ( 1.00,-0.62)--(B);

  \draw[thick,dash pattern=on 5pt off 3pt] (-0.62,-0.74)--(-0.62,0.74);
\end{tikzpicture}\,
\right)^{(1)}_{n,\ell}
\equiv
a^{(2)(1)}_{n,\ell} .
\label{diag28}
\end{equation}

\paragraph{Product rule.} Now we want to consider two-particle cuts which go through internal propagators. Such a cut splits the diagram into two subdiagrams, which are glued along the same two-particle state. We denote this gluing operation by $\otimes_2$.

We define the product rule as:
\begin{equation}\label{cutprod1}
    X\otimes_2 Y \equiv \frac{XY}{a^{(0)}_{n,\ell}} . 
\end{equation}
The division by $a^{(0)}_{n,\ell}$ removes the extra free OPE coefficient which is counted twice when the two cut subdiagrams are multiplied.

As a simple check, applying this rule to the disconnected two-particle cut gives $a^{(0)}_{n,\ell}\otimes_2 a^{(0)}_{n,\ell}=a^{(0)}_{n,\ell}$, as expected.

A diagram split into $f$ subdiagrams carries a label $\vec \nu=(\nu_1,\ldots,\nu_{f})$, where each entry specifies which OPE data identification is used for the corresponding subdiagram. Here $\nu_i\in\{0,1\}$. We keep only the choices satisfying $\sum_i \nu_i\leq 1$. Contributions with $\sum_i\nu_i>1$ can be absorbed into higher-order OPE data associated with diagrams with fewer cuts.

As a simple example, consider the case of one-loop bubble diagram, in which its internal two-particle cut factorizes into two tree-level cut diagrams:
\begin{equation}
\left(
\,\begin{tikzpicture}[
    baseline=-0.5ex,
    scale=0.75,
    line cap=round,
    line join=round
]
  \draw[thick] (0,0) circle (0.48);

  \draw[thick] (-1.25, 0.62)--(-0.48,0);
  \draw[thick] (-1.25,-0.62)--(-0.48,0);
  \draw[thick] ( 0.48,0)--(1.25, 0.62);
  \draw[thick] ( 0.48,0)--(1.25,-0.62);

  \draw[thick,dash pattern=on 5pt off 3pt] (0,-0.68)--(0,0.68);
\end{tikzpicture}\,
\right)^{(\nu_1,\nu_2)}_{n,\ell}
=
\left(
\,\begin{tikzpicture}[
    baseline=-0.5ex,
    scale=0.55,
    line cap=round,
    line join=round
]
  \draw[thick] (-1.00, 0.55)--(1.00,-0.55);
  \draw[thick] (-1.00,-0.55)--(1.00, 0.55);
  \draw[thick,dash pattern=on 5pt off 3pt] (0.35,-0.70)--(0.35,0.70);
\end{tikzpicture}\,
\right)^{(\nu_1)}_{n,\ell}
\otimes_2
\left(
\,\begin{tikzpicture}[
    baseline=-0.5ex,
    scale=0.55,
    line cap=round,
    line join=round
]
  \draw[thick] (-1.00, 0.55)--(1.00,-0.55);
  \draw[thick] (-1.00,-0.55)--(1.00, 0.55);
  \draw[thick,dash pattern=on 5pt off 3pt] (-0.35,-0.70)--(-0.35,0.70);
\end{tikzpicture}\,
\right)^{(\nu_2)}_{n,\ell}.
\label{diag29}
\end{equation}

For $\vec \nu=(0,0)$, this gives
\begin{equation}
\left(
\,\begin{tikzpicture}[
    baseline=-0.5ex,
    scale=0.75,
    line cap=round,
    line join=round
]
  \draw[thick] (0,0) circle (0.48);

  \draw[thick] (-1.25, 0.62)--(-0.48,0);
  \draw[thick] (-1.25,-0.62)--(-0.48,0);
  \draw[thick] ( 0.48,0)--(1.25, 0.62);
  \draw[thick] ( 0.48,0)--(1.25,-0.62);

  \draw[thick,dash pattern=on 5pt off 3pt] (0,-0.68)--(0,0.68);
\end{tikzpicture}\,
\right)^{(0,0)}_{n,\ell}
=
\left(a^{(0)}_{n,\ell}\gamma^{(1)(1)}_{n,\ell}\right)
\otimes_2
\left(a^{(0)}_{n,\ell}\gamma^{(1)(1)}_{n,\ell}\right)
=
a^{(0)}_{n,\ell}
\left(\gamma^{(1)(1)}_{n,\ell}\right)^2 .
\label{diag30}
\end{equation}

Similarly, for a $(r-1)$-loop bubble diagram with maximal internal cuts we have
\begin{equation}
\begin{aligned}
\left(
\,\begin{tikzpicture}[
    baseline=-0.5ex,
    scale=0.72,
    line cap=round,
    line join=round
]
  \def\r{0.48}

  \coordinate (B1) at (-1.05,0);
  \coordinate (B2) at (-0.09,0);
  \coordinate (B3) at ( 1.55,0);

  \draw[thick] (-1.95, 0.62)--(-1.53,0);
  \draw[thick] (-1.95,-0.62)--(-1.53,0);
  \draw[thick] ( 2.03,0)--(2.45, 0.62);
  \draw[thick] ( 2.03,0)--(2.45,-0.62);

  \draw[thick] (B1) circle (\r);
  \draw[thick] (B2) circle (\r);
  \draw[thick] (B3) circle (\r);

  \node at (0.75,0) {\large $\cdots$};

  \draw[thick,dash pattern=on 5pt off 3pt] (-1.05,-0.68)--(-1.05,0.68);
  \draw[thick,dash pattern=on 5pt off 3pt] (-0.09,-0.68)--(-0.09,0.68);
  \draw[thick,dash pattern=on 5pt off 3pt] ( 1.55,-0.68)--( 1.55,0.68);
\end{tikzpicture}\,
\right)^{(0,\ldots,0)}_{n,\ell}
&=
\underbrace{
\left(a^{(0)}_{n,\ell}\gamma^{(1)(1)}_{n,\ell}\right)
\otimes_2
\cdots
\otimes_2
\left(a^{(0)}_{n,\ell}\gamma^{(1)(1)}_{n,\ell}\right)
}_{r\ \text{factors}}
\\[0.4em]
&=
a^{(0)}_{n,\ell}
\left(\gamma^{(1)(1)}_{n,\ell}\right)^r .
\end{aligned}
\label{diag31}
\end{equation}
This is exactly the maximal-cut contribution studied in \cite{BubblesAdS}. This term, through crossing and inversion formula, is related to an external two-particle cut in the crossed channel. Namely,
\begin{equation}
\left(
\,\begin{tikzpicture}[
    baseline=-0.5ex,
    scale=0.82,
    line cap=round,
    line join=round
]
  \draw[thick] (-2.05, 0.72)--(-1.34, 0.20);
  \draw[thick] (-2.05,-0.72)--(-1.34,-0.20);

  \draw[thick] (1.10, 0.20)--(1.85, 0.72);
  \draw[thick] (1.10,-0.20)--(1.85,-0.72);

  \draw[thick] (-1.10,0) circle (0.30);
  \draw[thick] (-0.55,0) circle (0.30);
  \draw[thick] ( 0.85,0) circle (0.30);

  \node at (0.10,0) {$\cdots$};

  \draw[thick,dash pattern=on 5pt off 3pt] (-1.95,0.62)--(1.75,0.62);
\end{tikzpicture}\,
\right)^{(0)}_{n,\ell}
=
a^{(0)}_{n,\ell}\,\gamma^{(r)(1)}_{n,\ell}\, .
\label{diag32}
\end{equation}

\paragraph{Circled diagrams and crossing.} As we mentioned earlier, in our construction, a diagram inside parentheses denotes the OPE data factor multiplying a conformal block, while the corresponding circled diagram denotes its contribution to the correlator after summing over the exchanged double-trace operators.

Let $\Gamma_c$ be a diagram with a given two-particle cut denoted by $c$. We denote the corresponding OPE data factor by
\begin{equation}
    \left(\Gamma_c\right)^{\vec \nu}_{n,\ell}.
\end{equation}
Here $\vec\nu$ specifies the OPE data assignment associated with the subdiagrams produced by the cuts. Schematically, this factor is a product of OPE coefficients and anomalous dimensions of the form
\begin{equation}    
    \left(\Gamma_c\right)^{\vec \nu}_{n,\ell}  \sim a^{(s)}_{n,\ell}\prod_{m\geq 1} \left(\gamma^{(m)}_{n,\ell}\right)^{\omega_m}.
\end{equation}
The integer $\omega_m$ counts how many factors of the order $1/N^{2m}$ anomalous dimension appear in this term, and $s$ denotes the order of the OPE coefficient. Both $s$ and the $\omega_m$ are fixed by the diagram and by the choice of $\vec \nu$.

The corresponding contribution to the correlator is obtained by inserting this OPE data into the conformal-block expansion. Therefore, the associated circled diagram is defined by

\begin{equation} 
    \circdiag{\Gamma_c}^{\vec \nu} = \sum_{n,\ell} u^{\Delta_\CO+n} \left(\Gamma_c\right)^{\vec \nu}_{n,\ell} \frac{ \left(\log u+\partial_n\right)^{\Omega}}{\prod_{m\geq 1} 2^{\omega_m}\,\omega_m! } \tilde g_{n,\ell}(u,v) ,
\end{equation}
where $    \Omega=\sum_{m\geq 1}\omega_m$.

We can also sum over different $\vec\nu$ assignments and define
\begin{equation}
    \circdiag{\Gamma_c} \equiv \sum_{\vec \nu\in \mathcal N(\Gamma_c)} \circdiag{\Gamma_c}^{\vec \nu} .
\end{equation}
Here $\mathcal N(\Gamma_c)$ denotes the set of inequivalent allowed assignments. assignments which give exactly the same OPE data factor, are identified and should not be counted separately.

The full correlator at order $1/N^{2k}$ can then be organized as a sum over cut diagrams,
\begin{equation} 
    \mathcal G^{(k)}(u,v) = \sum_{\Gamma\in\mathcal D^{(k)}} \sum_{c\in \mathcal C_{\rm v}(\Gamma)} \frac{\mathfrak s_{\rm v}(\Gamma)}{2}\,\circdiag{\Gamma_c}\, .
\end{equation}
Here $\mathcal D^{(k)}$ denotes the set of connected diagrams contributing at order $1/N^{2k}$, and $\mathcal C_{\rm v}(\Gamma)$ denotes the set of inequivalent allowed vertical cuttings of a fixed diagram $\Gamma$. If two cuts give the same contribution to the conformal-block expansion, we keep only one representative. The factor $\mathfrak s_{\rm v}(\Gamma)$ is the dash-cut symmetry factor defined in \eqref{diag15}, evaluated for the vertical pairing. Therefore $\mathfrak s_{\rm v}(\Gamma)/2$ removes the overcounting from diagrams related only by this pairing exchange.

Crossing maps vertical cuts to horizontal cuts. Therefore the usual crossing equation for the full correlator can be written diagrammatically as
\begin{equation} \label{diag33}
    \sum_{\Gamma\in\mathcal D^{(k)}} \sum_{c\in \mathcal C_{\rm v}(\Gamma)}  \frac{\mathfrak s_{\rm v}(\Gamma)}{2}\, \circdiag{\Gamma_c}
    = \sum_{\Gamma\in\mathcal D^{(k)}} \sum_{c\in \mathcal C_{\rm h}(\Gamma)} \frac{\mathfrak s_{\rm h}(\Gamma)}{2}\, \circdiag{\Gamma_c}\, .
\end{equation}
Here $\mathcal C_{\rm h}(\Gamma)$ is defined analogously as the set of inequivalent allowed horizontal cuttings, and $\mathfrak s_{\rm h}(\Gamma)$ is the same dash-cut symmetry factor evaluated for the horizontal pairing.

\eqref{diag33} is simply the standard crossing relation rewritten in terms of cut diagrams. The advantage of the notation is that it allows us to ask a sharper question, whether crossing can be imposed on smaller classes of diagrams, rather than only on the full correlator.

With the diagrammatic representation, we can write a subset of correlator corresponding to a subset of diagrams which are closed under crossing.  For identical external particles one can take a fixed topology and sum over its channel realizations to obtain a crossing-symmetric combination. Therefore we can write
\begin{equation}  \label{diag34}
    \mathcal G^{(k)}_{[\Gamma]}(u,v) = \sum_{\Gamma'\in[\Gamma]} \sum_{c\in \mathcal C_{\rm v}(\Gamma')} \frac{\mathfrak s_{\rm v}(\Gamma')}{2}\, \circdiag{\Gamma'_c}\, .
\end{equation}
where $[\Gamma]$ denotes the crossing orbit of a diagram $\Gamma$, namely the set of diagrams obtained from $\Gamma$ by permuting the external points. If this orbit is closed under crossing, then we expect a relation of the form
\begin{equation} \label{diag35}
    \sum_{\Gamma'\in[\Gamma]} \sum_{c\in \mathcal C_{\rm v}(\Gamma')}\frac{\mathfrak s_{\rm v}(\Gamma')}{2}\,\circdiag{\Gamma'_c}
    =\sum_{\Gamma'\in[\Gamma]}\sum_{c\in \mathcal C_{\rm h}(\Gamma')} \frac{\mathfrak s_{\rm h}(\Gamma')}{2}\, \circdiag{\Gamma'_c}\, .
\end{equation}

At tree level, this gives two independent crossing equations for the contact and exchange topologies, which can be diagrammatically represented as
\begin{equation}
\begin{tikzpicture}[baseline=-0.5ex,scale=0.48,line cap=round,line join=round]
  \draw[thick] (0,0) circle (1.25);

  \draw (-0.82, 0.82)--( 0.82,-0.82);
  \draw (-0.82,-0.82)--( 0.82, 0.82);

  \draw[thick,dash pattern=on 5pt off 3pt] (-0.45,-0.90)--(-0.45,0.90);
\end{tikzpicture}
=
\begin{tikzpicture}[baseline=-0.5ex,scale=0.48,line cap=round,line join=round]
  \draw[thick] (0,0) circle (1.25);

  \draw (-0.82, 0.82)--( 0.82,-0.82);
  \draw (-0.82,-0.82)--( 0.82, 0.82);

  \draw[thick,dash pattern=on 5pt off 3pt] (-0.90,0.35)--(0.90,0.35);
\end{tikzpicture}\, .
\label{diag36}
\end{equation}

\begin{equation}
\begin{tikzpicture}[baseline=-0.5ex,scale=0.48,line cap=round,line join=round]
  \draw[thick] (0,0) circle (1.25);

  \coordinate (L) at (-0.45,0);
  \coordinate (R) at ( 0.45,0);

  \draw (L)--(R);

  \draw (-1.05, 0.78)--(L);
  \draw (-1.05,-0.78)--(L);
  \draw (R)--(1.05, 0.78);
  \draw (R)--(1.05,-0.78);

  \draw[thick,dash pattern=on 5pt off 3pt] (-0.78,-0.75)--(-0.78,0.75);
\end{tikzpicture}
\!+\!
\begin{tikzpicture}[baseline=-0.5ex,scale=0.48,line cap=round,line join=round]
  \draw[thick] (0,0) circle (1.25);

  \coordinate (L) at (-0.45,0);
  \coordinate (R) at ( 0.45,0);

  \draw (L)--(R);

  \draw (-1.05, 0.78)--(L);
  \draw (-1.05,-0.78)--(L);
  \draw (R)--(1.05, 0.78);
  \draw (R)--(1.05,-0.78);

  \draw[thick,dash pattern=on 5pt off 3pt] (0,-0.75)--(0,0.75);
\end{tikzpicture}
\!+\!
\begin{tikzpicture}[baseline=-0.5ex,scale=0.48,line cap=round,line join=round]
  \draw[thick] (0,0) circle (1.25);

  \coordinate (T) at (0, 0.45);
  \coordinate (B) at (0,-0.45);

  \draw  (T)--(B);

  \draw (-0.75, 1.05)--(T);
  \draw ( 0.75, 1.05)--(T);
  \draw (-0.75,-1.05)--(B);
  \draw ( 0.75,-1.05)--(B);

  \draw[thick,dash pattern=on 5pt off 3pt] (-0.38,-0.95)--(-0.38,0.95);
\end{tikzpicture}
=
\begin{tikzpicture}[baseline=-0.5ex,scale=0.48,line cap=round,line join=round]
  \draw[thick] (0,0) circle (1.25);

  \coordinate (L) at (-0.45,0);
  \coordinate (R) at ( 0.45,0);

  \draw (L)--(R);

  \draw (-1.05, 0.78)--(L);
  \draw (-1.05,-0.78)--(L);
  \draw (R)--(1.05, 0.78);
  \draw (R)--(1.05,-0.78);

  \draw[thick,dash pattern=on 5pt off 3pt] (-0.90,0.35)--(0.90,0.35);
\end{tikzpicture}
\!+\!
\begin{tikzpicture}[baseline=-0.5ex,scale=0.48,line cap=round,line join=round]
  \draw[thick] (0,0) circle (1.25);

  \coordinate (T) at (0, 0.45);
  \coordinate (B) at (0,-0.45);

  \draw (T)--(B);

  \draw (-0.75, 1.05)--(T);
  \draw ( 0.75, 1.05)--(T);
  \draw (-0.75,-1.05)--(B);
  \draw ( 0.75,-1.05)--(B);

  \draw[thick,dash pattern=on 5pt off 3pt] (-0.90,0.65)--(0.90,0.65);
\end{tikzpicture}
\!+\!
\begin{tikzpicture}[baseline=-0.5ex,scale=0.48,line cap=round,line join=round]
  \draw[thick] (0,0) circle (1.25);

  \coordinate (T) at (0, 0.45);
  \coordinate (B) at (0,-0.45);

  \draw (T)--(B);

  \draw (-0.75, 1.05)--(T);
  \draw ( 0.75, 1.05)--(T);
  \draw (-0.75,-1.05)--(B);
  \draw ( 0.75,-1.05)--(B);

  \draw[thick,dash pattern=on 5pt off 3pt] (-0.90,0)--(0.90,0);
\end{tikzpicture}\, .
\label{diag37}
\end{equation}
We know that \eqref{diag36} and \eqref{diag37} hold true, for example from \cite{Alday:2017gde}. However, \eqref{diag35} can generate more interesting crossing relations at higher orders. For example, at two loops in $\phi^4$ theory, the 2-bubble topology is not the only topology contributing to the correlator. Nevertheless, its crossing orbit forms a crossing-closed class by itself. We can therefore write
\begin{equation}
\resizebox{\textwidth}{!}{$
\begin{tikzpicture}[baseline=-0.5ex,scale=0.50,line cap=round,line join=round]
  \draw[thick] (0,0) circle (1.25);

  \draw (-0.28,0) circle (0.28);
  \draw ( 0.28,0) circle (0.28);

  \draw (-1.15, 0.45)--(-0.56,0);
  \draw (-1.15,-0.45)--(-0.56,0);
  \draw ( 0.56,0)--(1.15, 0.45);
  \draw ( 0.56,0)--(1.15,-0.45);

  \draw[thick,dash pattern=on 5pt off 3pt] (-0.88,-0.60)--(-0.88,0.60);
\end{tikzpicture}
\!+\!
\begin{tikzpicture}[baseline=-0.5ex,scale=0.50,line cap=round,line join=round]
  \draw[thick] (0,0) circle (1.25);

  \draw (-0.28,0) circle (0.28);
  \draw ( 0.28,0) circle (0.28);

  \draw (-1.15, 0.45)--(-0.56,0);
  \draw (-1.15,-0.45)--(-0.56,0);
  \draw ( 0.56,0)--(1.15, 0.45);
  \draw ( 0.56,0)--(1.15,-0.45);

  \draw[thick,dash pattern=on 5pt off 3pt] (-0.28,-0.62)--(-0.28,0.62);
\end{tikzpicture}
\!+\!
\begin{tikzpicture}[baseline=-0.5ex,scale=0.50,line cap=round,line join=round]
  \draw[thick] (0,0) circle (1.25);

  \draw (-0.28,0) circle (0.28);
  \draw ( 0.28,0) circle (0.28);

  \draw (-1.15, 0.45)--(-0.56,0);
  \draw (-1.15,-0.45)--(-0.56,0);
  \draw ( 0.56,0)--(1.15, 0.45);
  \draw ( 0.56,0)--(1.15,-0.45);

  \draw[thick,dash pattern=on 5pt off 3pt] (-0.28,-0.62)--(-0.28,0.62);
  \draw[thick,dash pattern=on 5pt off 3pt] ( 0.28,-0.62)--( 0.28,0.62);
\end{tikzpicture}
\!+\!
\begin{tikzpicture}[baseline=-0.5ex,scale=0.50,line cap=round,line join=round]
  \draw[thick] (0,0) circle (1.25);

  \draw (0, 0.30) circle (0.28);
  \draw (0,-0.30) circle (0.28);

  \draw (-0.75, 1.00)--(0, 0.58);
  \draw ( 0.75, 1.00)--(0, 0.58);
  \draw (-0.75,-1.00)--(0,-0.58);
  \draw ( 0.75,-1.00)--(0,-0.58);

  \draw[thick,dash pattern=on 5pt off 3pt] (-0.45,-0.95)--(-0.45,0.95);
\end{tikzpicture}
=
\begin{tikzpicture}[baseline=-0.5ex,scale=0.50,line cap=round,line join=round]
  \draw[thick] (0,0) circle (1.25);

  \draw (0, 0.30) circle (0.28);
  \draw (0,-0.30) circle (0.28);

  \draw (-0.75, 1.00)--(0, 0.58);
  \draw ( 0.75, 1.00)--(0, 0.58);
  \draw (-0.75,-1.00)--(0,-0.58);
  \draw ( 0.75,-1.00)--(0,-0.58);

  \draw[thick,dash pattern=on 5pt off 3pt] (-0.80,0.78)--(0.80,0.78);
\end{tikzpicture}
\!+\!
\begin{tikzpicture}[baseline=-0.5ex,scale=0.50,line cap=round,line join=round]
  \draw[thick] (0,0) circle (1.25);

  \draw (0, 0.30) circle (0.28);
  \draw (0,-0.30) circle (0.28);

  \draw (-0.75, 1.00)--(0, 0.58);
  \draw ( 0.75, 1.00)--(0, 0.58);
  \draw (-0.75,-1.00)--(0,-0.58);
  \draw ( 0.75,-1.00)--(0,-0.58);

  \draw[thick,dash pattern=on 5pt off 3pt] (-0.80,0.30)--(0.80,0.30);
\end{tikzpicture}
\!+\!
\begin{tikzpicture}[baseline=-0.5ex,scale=0.50,line cap=round,line join=round]
  \draw[thick] (0,0) circle (1.25);

  \draw (0, 0.30) circle (0.28);
  \draw (0,-0.30) circle (0.28);

  \draw (-0.75, 1.00)--(0, 0.58);
  \draw ( 0.75, 1.00)--(0, 0.58);
  \draw (-0.75,-1.00)--(0,-0.58);
  \draw ( 0.75,-1.00)--(0,-0.58);

  \draw[thick,dash pattern=on 5pt off 3pt] (-0.80, 0.30)--(0.80, 0.30);
  \draw[thick,dash pattern=on 5pt off 3pt] (-0.80,-0.30)--(0.80,-0.30);
\end{tikzpicture}
\!+\!
\begin{tikzpicture}[baseline=-0.5ex,scale=0.50,line cap=round,line join=round]
  \draw[thick] (0,0) circle (1.25);

  \draw (-0.28,0) circle (0.28);
  \draw ( 0.28,0) circle (0.28);

  \draw (-1.05, 0.75)--(-0.56,0);
  \draw (-1.05,-0.75)--(-0.56,0);
  \draw ( 0.56,0)--(1.05, 0.75);
  \draw ( 0.56,0)--(1.05,-0.75);

  \draw[thick,dash pattern=on 5pt off 3pt] (-0.75,0.50)--(0.75,0.50);
\end{tikzpicture}\, .
$}
\end{equation}
\subsection{Crossing and  inversion formula}

In the previous subsection we organized crossing at the level of crossing-closed orbits. For a diagram topology $\Gamma$, we wrote the corresponding orbit contribution after summing over all channel realizations $\Gamma'\in[\Gamma]$ as \eqref{diag34}. We now want to refine this organization by keeping a single channel realization $\Gamma'$. This cannot be achieved directly using dash-cuts, since dash-cuts can mix different channel realizations. Therefore, to isolate a fixed $\Gamma'$, we use dot-cuts.

We write
\begin{equation}\label{diag38}
    \mathcal G_{\Gamma'}(u,v) \equiv \sum_{\dot c\in \dot{\mathcal C}_{\rm v}(\Gamma')} \circdiag{\Gamma'_{\dot c}}\, ,
\end{equation}
where $\dot{\mathcal C}_{\rm v}(\Gamma')$ denotes the set of vertical dot-cuts on $\Gamma'$.

Equation~\eqref{diag38} should be understood as an auxiliary representation. The left-hand side is the ordinary Witten diagram associated with the fixed channel realization $\Gamma'$. But the right-hand side is not written in the meaningful dash-cut dictionary of the conformal-block expansion. We remind that individual dot-cuts do not directly correspond to the OPE data. However, this representation is still useful because the dot-cuts keep track of the pairings explicitly and can later be recombined into dash-cuts.

We can now use this auxiliary representation to write a refined crossing relation. Since $\mathcal G_{\Gamma'}(u,v)$ is the ordinary Witten diagram in a fixed channel, crossing acts on it in the usual way. In the dot-cut representation this gives
\begin{equation}   \label{diag39}
    \sum_{\dot c\in \dot{\mathcal C}_{\rm v}(\Gamma')} \circdiag{\Gamma'_{\dot c}} = \sum_{\dot c\in \dot{\mathcal C}_{\rm h}(\Gamma')} \circdiag{\Gamma'_{\dot c}}\, .
\end{equation}
Here $\dot{\mathcal C}_{\rm h}(\Gamma')$ denotes the set of horizontal dot-cuts. This equation should still be understood as an auxiliary statement, since the dot-cut terms do not directly translate into the conformal-block expansion.

However, in many cases the dot-cut equation can be rewritten by replacing the dot-cuts with the meaningful dash-cuts, up to external horizontal cuts. For example, for the exchange diagram one obtains relations of the schematic form:
\begin{equation}
\begin{tikzpicture}[baseline=-0.5ex,scale=0.65,line cap=round,line join=round]
  \draw[thick] (0,0) circle (1.10);

  \coordinate (T) at (0, 0.35);
  \coordinate (B) at (0,-0.35);

  \draw (T)--(B);

  \draw (-0.55, 0.92)--(T);
  \draw ( 0.55, 0.92)--(T);
  \draw (-0.55,-0.92)--(B);
  \draw ( 0.55,-0.92)--(B);

  \draw[thick,dash pattern=on 5pt off 3pt] (-0.40,0.82)--(-0.40,-0.82);
\end{tikzpicture}
=
\begin{tikzpicture}[baseline=-0.5ex,scale=0.65,line cap=round,line join=round]
  \draw[thick] (0,0) circle (1.10);

  \coordinate (T) at (0, 0.35);
  \coordinate (B) at (0,-0.35);

  \draw (T)--(B);

  \draw (-0.55, 0.92)--(T);
  \draw ( 0.55, 0.92)--(T);
  \draw (-0.55,-0.92)--(B);
  \draw ( 0.55,-0.92)--(B);

  \draw[thick,dash pattern=on 5pt off 3pt] (-0.68,0)--(0.68,0);
\end{tikzpicture}
+
\bigl[\text{external horizontal cuts}\bigr] .
\label{diag37int}
\end{equation}

\begin{equation}
\begin{tikzpicture}[baseline=-0.5ex,scale=0.65,line cap=round,line join=round]
  \draw[thick] (0,0) circle (1.10);

  \coordinate (L) at (-0.35,0);
  \coordinate (R) at ( 0.35,0);
  \draw (L)--(R);
  \draw (-0.92, 0.55)--(L);
  \draw (-0.92,-0.55)--(L);
  \draw (R)--(0.92, 0.55);
  \draw (R)--(0.92,-0.55);

  \draw[thick,dash pattern=on 5pt off 3pt] (-0.62,-0.82)--(-0.62,0.82);
\end{tikzpicture}
\;+\;
\begin{tikzpicture}[baseline=-0.5ex,scale=0.65,line cap=round,line join=round]
  \draw[thick] (0,0) circle (1.10);

  \coordinate (L) at (-0.35,0);
  \coordinate (R) at ( 0.35,0);

  \draw  (L)--(R);

  \draw (-0.92, 0.55)--(L);
  \draw (-0.92,-0.55)--(L);
  \draw (R)--(0.92, 0.55);
  \draw (R)--(0.92,-0.55);

  \draw[thick,dash pattern=on 5pt off 3pt] (0,-0.82)--(0,0.82);
\end{tikzpicture}
=
\bigl[\text{external horizontal cuts}\bigr] .
\label{diag37ext}
\end{equation}

Diagrams with external horizontal cuts have vanishing double discontinuity in the crossed channel. Hence they can be ignored when applying the inversion formula. Therefore, to read off the OPE data associated with a given vertical cut, we apply the inversion formula to the possible internal horizontal cuts.

If a diagram does not admit any internal horizontal cut, as in \eqref{diag37ext}, then the corresponding OPE data can only have finite support in spin. On the other hand, for the case \eqref{diag37int}, we can write
\begin{equation}
\left(
\begin{tikzpicture}[baseline=-0.5ex,scale=0.65,line cap=round,line join=round]
  \coordinate (T) at (0, 0.35);
  \coordinate (B) at (0,-0.35);

  \draw (T)--(B);
  \draw (-0.55, 0.92)--(T);
  \draw ( 0.55, 0.92)--(T);
  \draw (-0.55,-0.92)--(B);
  \draw ( 0.55,-0.92)--(B);
  \draw[thick,dash pattern=on 5pt off 3pt] (-0.40,0.82)--(-0.40,-0.82);
\end{tikzpicture}
\right)^{(0)}_{n,\ell}
=
\Bigg(\mathrm{LIF}\!\left[
\begin{tikzpicture}[baseline=-0.5ex,scale=0.65,line cap=round,line join=round]
  \draw[thick] (0,0) circle (1.10);

  \coordinate (T) at (0, 0.35);
  \coordinate (B) at (0,-0.35);

  \draw (T)--(B);
  \draw (-0.55, 0.92)--(T);
  \draw ( 0.55, 0.92)--(T);
  \draw (-0.55,-0.92)--(B);
  \draw ( 0.55,-0.92)--(B);

  \draw[thick,dash pattern=on 5pt off 3pt] (-0.68,0)--(0.68,0);
\end{tikzpicture}
\right]\Bigg)_{[\CO\CO]_{n,\ell}}^{(2)}.
\label{diag40}
\end{equation}
Where
\begin{equation*}
   \Big(\mathrm{LIF}[\mathcal X]\Big)^{(i)}_{[\CO\CO]_{n,\ell}} 
\end{equation*}

denotes the coefficient of the $i$-th order pole of the inversion function obtained from $\mathcal X\subset \mathcal{G}(u,v)$,  at the double-trace position $\Delta = 2\Delta_\CO+2n+\ell $

Using \eqref{diag40} we can find the known asymptotic solution. For $\Delta_\CO=2$ this gives

\begin{equation}  \label{gam1(2)}
    \gamma^{(1)(2)}_{n,\ell} = -\frac{ 2a_{2,0} }{ (\ell+1)(\ell+2n+2) } .
\end{equation}

This fixes the part of the exchange contribution associated with the internal horizontal cut. To determine the remaining part, we use the full crossing relation for the exchange topology, written in the form \eqref{diag37}. This gives the finite-spin contribution
\begin{equation}\label{gam1(3)}
\gamma^{(1)(3)}_{n,\ell} =\delta_{\ell,0}\, \frac{1}{2}\, \frac{(n+1)}{(2n+1)(2n+3)}\,a_{2,0}.
\end{equation}
The support at $\ell=0$ is consistent with the fact that this contribution is not associated with an internal horizontal cut.

This split of the tree-level exchange anomalous dimension into an asymptotic piece and a finite-spin piece was already observed in~\cite{Alday:2017gde}. The new point here is that this two pieces admit distinct diagrammatic representations, as in \eqref{diag26} and \eqref{diag27}.

\subsubsection{One-loop check}\label{lastsub}

As an application, and also as a check of the diagrammatic interpretation above, we can build one-loop objects by gluing the tree-level building blocks. We have three tree-level pieces,
\begin{equation}
\gamma^{(1)(1)}_{n,\ell},\qquad \gamma^{(1)(2)}_{n,\ell},\qquad \gamma^{(1)(3)}_{n,\ell},
\end{equation}
with their diagrammatic representation given in \eqref{diag25}, \eqref{diag26}, and \eqref{diag27}.

By merging these tree-level data we can obtain different one-loop diagrams. Namely
\begin{alignat}{2}
\Gamma_1
&=
\begin{tikzpicture}[baseline=-0.5ex,scale=0.55,line cap=round,line join=round]
  \draw[thick] (0,0) ellipse (0.22 and 0.38);
  \draw[thick] (-0.70, 0.85)--(0, 0.32);
  \draw[thick] ( 0.70, 0.85)--(0, 0.32);
  \draw[thick] (-0.70,-0.85)--(0,-0.32);
  \draw[thick] ( 0.70,-0.85)--(0,-0.32);
\end{tikzpicture}
,
\qquad
&\left(
\begin{tikzpicture}[baseline=-0.5ex,scale=0.55,line cap=round,line join=round]
  \draw[thick] (0,0) ellipse (0.22 and 0.38);
  \draw[thick] (-0.70, 0.85)--(0, 0.32);
  \draw[thick] ( 0.70, 0.85)--(0, 0.32);
  \draw[thick] (-0.70,-0.85)--(0,-0.32);
  \draw[thick] ( 0.70,-0.85)--(0,-0.32);
  \draw[thick,dash pattern=on 5pt off 3pt] (-0.58,0)--(0.58,0);
\end{tikzpicture}
\right)^{(0)}_{n,\ell}
&=
a^{(0)}_{n,\ell}
\left(\gamma^{(1)(1)}_{n,\ell}\right)^2
\\
\Gamma_2
&=
\begin{tikzpicture}[baseline=-0.5ex,scale=0.55,line cap=round,line join=round]
  \coordinate (T) at (0,0.48);
  \coordinate (L) at (-0.58,-0.45);
  \coordinate (R) at ( 0.58,-0.45);
  \draw[thick] (T)--(L)--(R)--cycle;
  \draw[thick] (-0.70, 0.88)--(T);
  \draw[thick] ( 0.70, 0.88)--(T);
  \draw[thick] (-0.95,-0.95)--(L);
  \draw[thick] ( 0.95,-0.95)--(R);
\end{tikzpicture}
,
\qquad
&\left(
\begin{tikzpicture}[baseline=-0.5ex,scale=0.55,line cap=round,line join=round]
  \coordinate (T) at (0,0.48);
  \coordinate (L) at (-0.58,-0.45);
  \coordinate (R) at ( 0.58,-0.45);
  \draw[thick] (T)--(L)--(R)--cycle;
  \draw[thick] (-0.70, 0.88)--(T);
  \draw[thick] ( 0.70, 0.88)--(T);
  \draw[thick] (-0.95,-0.95)--(L);
  \draw[thick] ( 0.95,-0.95)--(R);
  \draw[thick,dash pattern=on 5pt off 3pt] (-0.48,-0.05)--(0.48,-0.05);
\end{tikzpicture}
\right)^{(0)}_{n,\ell}
&=
a^{(0)}_{n,\ell}
\gamma^{(1)(1)}_{n,\ell}
\gamma^{(1)(2)}_{n,\ell}
\\
\Gamma_3
&=
\begin{tikzpicture}[baseline=-0.5ex,scale=0.55,line cap=round,line join=round]
  \draw[thick] (0,0.18) ellipse (0.22 and 0.38);
  \draw[thick] (-0.70, 1.02)--(0, 0.50);
  \draw[thick] ( 0.70, 1.02)--(0, 0.50);
  \draw[thick] (0,-0.20)--(0,-0.75);
  \draw[thick] (-0.70,-1.25)--(0,-0.75);
  \draw[thick] ( 0.70,-1.25)--(0,-0.75);
\end{tikzpicture}
,
\qquad
&\left(
\begin{tikzpicture}[baseline=-0.5ex,scale=0.55,line cap=round,line join=round]
  \draw[thick] (0,0.18) ellipse (0.22 and 0.38);
  \draw[thick] (-0.70, 1.02)--(0, 0.50);
  \draw[thick] ( 0.70, 1.02)--(0, 0.50);
  \draw[thick] (0,-0.20)--(0,-0.75);
  \draw[thick] (-0.70,-1.25)--(0,-0.75);
  \draw[thick] ( 0.70,-1.25)--(0,-0.75);
  \draw[thick,dash pattern=on 5pt off 3pt] (-0.58,0.18)--(0.58,0.18);
\end{tikzpicture}
\right)^{(0)}_{n,\ell}
&=
a^{(0)}_{n,\ell}
\gamma^{(1)(1)}_{n,\ell}
\gamma^{(1)(3)}_{n,\ell}
\\
\Gamma_4
&=
\begin{tikzpicture}[baseline=-0.5ex,scale=0.55,line cap=round,line join=round]
  \coordinate (V) at (0,0.95);
  \coordinate (T) at (0,0.35);
  \coordinate (L) at (-0.58,-0.55);
  \coordinate (R) at ( 0.58,-0.55);
  \draw[thick] (T)--(L)--(R)--cycle;
  \draw[thick] (V)--(T);
  \draw[thick] (-0.70,1.45)--(V);
  \draw[thick] ( 0.70,1.45)--(V);
  \draw[thick] (-0.95,-1.05)--(L);
  \draw[thick] ( 0.95,-1.05)--(R);
\end{tikzpicture}
,
\qquad
&\left(
\begin{tikzpicture}[baseline=-0.5ex,scale=0.55,line cap=round,line join=round]
  \coordinate (V) at (0,0.95);
  \coordinate (T) at (0,0.35);
  \coordinate (L) at (-0.58,-0.55);
  \coordinate (R) at ( 0.58,-0.55);
  \draw[thick] (T)--(L)--(R)--cycle;
  \draw[thick] (V)--(T);
  \draw[thick] (-0.70,1.45)--(V);
  \draw[thick] ( 0.70,1.45)--(V);
  \draw[thick] (-0.95,-1.05)--(L);
  \draw[thick] ( 0.95,-1.05)--(R);
  \draw[thick,dash pattern=on 5pt off 3pt] (-0.48,-0.12)--(0.48,-0.12);
\end{tikzpicture}
\right)^{(0)}_{n,\ell}
&=
a^{(0)}_{n,\ell}
\gamma^{(1)(2)}_{n,\ell}
\gamma^{(1)(3)}_{n,\ell}
\\
\Gamma_5
&=
\begin{tikzpicture}[baseline=-0.5ex,scale=0.55,line cap=round,line join=round]
  \draw[thick] (0,0) ellipse (0.22 and 0.38);
  \draw[thick] (0, 0.38)--(0, 0.95);
  \draw[thick] (0,-0.38)--(0,-0.95);
  \draw[thick] (-0.70, 1.45)--(0, 0.95);
  \draw[thick] ( 0.70, 1.45)--(0, 0.95);
  \draw[thick] (-0.70,-1.45)--(0,-0.95);
  \draw[thick] ( 0.70,-1.45)--(0,-0.95);
\end{tikzpicture}
,
\qquad
&\left(
\begin{tikzpicture}[baseline=-0.5ex,scale=0.55,line cap=round,line join=round]
  \draw[thick] (0,0) ellipse (0.22 and 0.38);
  \draw[thick] (0, 0.38)--(0, 0.95);
  \draw[thick] (0,-0.38)--(0,-0.95);
  \draw[thick] (-0.70, 1.45)--(0, 0.95);
  \draw[thick] ( 0.70, 1.45)--(0, 0.95);
  \draw[thick] (-0.70,-1.45)--(0,-0.95);
  \draw[thick] ( 0.70,-1.45)--(0,-0.95);
  \draw[thick,dash pattern=on 5pt off 3pt] (-0.58,0)--(0.58,0);
\end{tikzpicture}
\right)^{(0)}_{n,\ell}
&=
a^{(0)}_{n,\ell}
\left(\gamma^{(1)(3)}_{n,\ell}\right)^2 .
\end{alignat}
We can use the discussion in the Appendix \ref{app2} to obtain the Mellin amplitudes associated to each of these diagrams.  The Mellin amplitude has the form \eqref{mmamp} and corresponding $R_m^{(1)}$ factors can be read off from \eqref{rramp}. So we find
\begin{equation}
\begin{alignedat}{2}
R_m^{(1)}(\Gamma_1)&= -\frac{9(m+1)(3m+4)(1)_m}{2(2m+3)\left(\frac32\right)_m},\qquad&R_m^{(1)}(\Gamma_2)&=\frac{6(1)_m}{\left(\frac32\right)_m}\,a_{2,0},\\[0.8em]R_m^{(1)}(\Gamma_3)&=-\frac{3(m+2)(1)_m}{2(2m+3)\left(\frac32\right)_m}\,a_{2,0},\qquad&R_m^{(1)}(\Gamma_4)&=\frac{2(m+1)(1)_m-\left(\frac32\right)_m}{(m+1)^2\left(\frac32\right)_m}\,(a_{2,0})^2,\\[0.8em]R_m^{(1)}(\Gamma_5)&=-\frac{(1)_m}{2(2m+3)\left(\frac32\right)_m}\,(a_{2,0})^2 .
\end{alignedat}
\end{equation}
Now inserting these data into the flat-space limit formula \eqref{flatlimit} we find the corresponding flat-space amplitudes
\begin{align}
\mathcal{A}(\Gamma_1 ) &=\frac{9}{8} \pi R_{\text{AdS}}^2  \sqrt{- T}, &\qquad \mathcal{A}(\Gamma_2 )  &= \frac{6 \pi }{\sqrt{- T}} a_{2,0}, \\ \mathcal{A}(\Gamma_3)  &= -\frac{3 \pi }{4 \sqrt{- T}} a_{2,0},&\qquad \mathcal{A}(\Gamma_4 )  &= -\frac{4 \pi }{R_{\text{AdS}}^2\left(- T\right)^{3/2}} (a_{2,0})^2, \\\mathcal{A}(\Gamma_5 )  &= \frac{\pi }{2 R_{\text{AdS}}^2\left(- T\right)^{3/2}} (a_{2,0})^2 .
\end{align}
This is exactly the corresponding flat-space amplitude in the massless limit and upon fixing
\begin{equation}
\lambda=\frac{48R_{\text{AdS}}\pi^2}{N^2},\qquad g^2=\frac{32\pi^2}{R_{\text{AdS}}N^2}a_{2,0} ,
\end{equation}
where $\lambda$ and $g$ are the quartic and cubic coupling constants. 

This shows consistency of associating $\gamma_{n,\ell}^{(1)(2)}$ and $\gamma_{n,\ell}^{(1)(3)}$ given in \eqref{gam1(2)} and \eqref{gam1(3)} with diagrams \eqref{diag26}, and \eqref{diag27}.

\subsection{Diagrammatic interpretation of the higher-trace analysis}

Finally, we can try to diagrammatically express the  calculation performed in Section~\ref{sec3}. The advantage of the diagrammatic representation is that it makes manifest, already at the level of cuts, which terms in the double-trace sector are related by crossing to the appearance of higher-trace operators.

In particular, for $\phi^4$ theory at three loops, the coefficient $b^{(0)}_{q,\ell}$ studied in Section~\ref{subopecoef} is associated with the vertical cut of the diagram shown in \eqref{b0lif}. Diagrammatically, what we did was to determine this OPE data by applying the Lorentzian inversion formula to the corresponding horizontal internal cut in the crossed channel,
\begin{equation} \label{b0lif}
b^{(0)}_{q,\ell} = \left(
\begin{tikzpicture}[baseline=-0.5ex,scale=0.9,line cap=round,line join=round]

  \draw[thick] (0,0) ellipse (0.42 and 0.52);

  \draw[thick]
    (-1.10, 0.82)
    .. controls (-0.70,0.55) and (-0.45,0.34) ..
    (-0.18,0.28)
    .. controls (0.20,0.20) and (0.55,0.42) ..
    (1.10,0.82);

  \draw[thick]
    (-1.10,-0.82)
    .. controls (-0.70,-0.55) and (-0.45,-0.34) ..
    (-0.18,-0.28)
    .. controls (0.20,-0.20) and (0.55,-0.42) ..
    (1.10,-0.82);

  \draw[thick,dash pattern=on 5pt off 3pt] (0,1.08)--(0,-1.08);

\end{tikzpicture}
\right)^{(0)}_{q,\ell}
=
\left(
\mathrm{LIF}\!\left[
\begin{tikzpicture}[baseline=-0.5ex,scale=0.9,line cap=round,line join=round]

  \draw[thick] (0,0) circle (1.12);

  \draw[thick] (0,0) ellipse (0.42 and 0.52);

  \draw[thick]
    (-0.82, 0.78)
    .. controls (-0.58,0.55) and (-0.42,0.34) ..
    (-0.18,0.28)
    .. controls (0.20,0.20) and (0.55,0.42) ..
    (0.82,0.78);

  \draw[thick]
    (-0.82,-0.78)
    .. controls (-0.58,-0.55) and (-0.42,-0.34) ..
    (-0.18,-0.28)
    .. controls (0.20,-0.20) and (0.55,-0.42) ..
    (0.82,-0.78);

  \draw[thick,dash pattern=on 5pt off 3pt] (-0.78,0)--(0.78,0);

\end{tikzpicture}
\right]
\right)^{(1)}_{[\CO^4]_{q,\ell}} .
\end{equation}

The important point is that this relation can be read directly from the diagrams, before carrying out the explicit spin sum. The horizontal internal cut on the right hand side is exactly the contribution whose crossed-channel double discontinuity produces the direct-channel data on the left-hand side. In this way, the diagrammatic representation tells us in advance which part of the double-trace sector is responsible, after crossing, for the higher-trace operators that appear in the direct channel.

Similarly, the calculation in section \ref{subanom} that obtained the anomalous dimension of this quadruple-trace family can be diagrammatically represented as
\begin{equation}
b^{(0)}_{q,\ell}\,\delta^{(1)}_{q,\ell}=\left(\mathrm{LIF}\!\left[
\begin{tikzpicture}[baseline=-0.5ex,scale=0.95,line cap=round,line join=round]

  \draw[thick] (0,0) circle (1.18);

  \coordinate (TL) at (-0.58, 0.36);
  \coordinate (TM) at ( 0.00, 0.36);
  \coordinate (TR) at ( 0.58, 0.36);
  \coordinate (BL) at (-0.42,-0.46);
  \coordinate (BR) at ( 0.42,-0.46);

  \draw[thick] (-0.92, 0.78)--(TL);
  \draw[thick] ( 0.92, 0.78)--(TR);
  \draw[thick] (-0.92,-0.78)--(BL);
  \draw[thick] ( 0.92,-0.78)--(BR);

  \draw[thick] (TL)--(BL);
  \draw[thick] (TR)--(BR);

  \draw[thick]
    (TL) .. controls (-0.46,0.60) and (-0.12,0.60) .. (TM);
  \draw[thick]
    (TL) .. controls (-0.46,0.12) and (-0.12,0.12) .. (TM);

  \draw[thick]
    (TM) .. controls (0.12,0.60) and (0.46,0.60) .. (TR);
  \draw[thick]
    (TM) .. controls (0.12,0.12) and (0.46,0.12) .. (TR);

  \draw[thick]
    (BL) .. controls (-0.25,-0.22) and (0.25,-0.22) .. (BR);
  \draw[thick]
    (BL) .. controls (-0.25,-0.70) and (0.25,-0.70) .. (BR);

  \draw[thick,dash pattern=on 5pt off 3pt] (-0.88,-0.05)--(0.88,-0.05);

\end{tikzpicture}
\right]
\right)^{(2)}_{[\CO^4]_{q,\ell}} .
\end{equation}

For the $\phi^3$ theory, the same idea applies to the ladder diagrams studied in Section~\ref{subphi3}. In that case the leading large-spin logarithm is obtained from the horizontal cuts of the ladder diagram. Diagrammatically
\begin{equation}
\left.
\left[
\begin{aligned}
\left(
\begin{tikzpicture}[baseline=-0.5ex,scale=0.92,line cap=round,line join=round]

  \def\xL{-0.38}
  \def\xR{ 0.38}

  \draw[thick] (-0.95, 1.65)--(\xL,1.25);
  \draw[thick] ( 0.95, 1.65)--(\xR,1.25);
  \draw[thick] (-0.95,-1.50)--(\xL,-1.10);
  \draw[thick] ( 0.95,-1.50)--(\xR,-1.10);

  \draw[thick] (\xL,1.25)--(\xL,0.35);
  \draw[thick] (\xR,1.25)--(\xR,0.35);
  \draw[thick] (\xL,1.25)--(\xR,1.25);
  \draw[thick] (\xL,0.80)--(\xR,0.80);
  \draw[thick] (\xL,0.35)--(\xR,0.35);

  \draw[thick] (\xL,0.35)--(\xL,0.12);
  \draw[thick] (\xR,0.35)--(\xR,0.12);

  \node at (0,-0.18) {$\vdots$};

  \draw[thick] (\xL,-0.55)--(\xL,-0.78);
  \draw[thick] (\xR,-0.55)--(\xR,-0.78);

  \draw[thick] (\xL,-0.78)--(\xL,-1.10);
  \draw[thick] (\xR,-0.78)--(\xR,-1.10);
  \draw[thick] (\xL,-0.78)--(\xR,-0.78);
  \draw[thick] (\xL,-1.10)--(\xR,-1.10);

  \draw[thick,dash pattern=on 5pt off 3pt] (0,1.45)--(0,-1.28);

\end{tikzpicture}
\right)^{(0)}_{r,\ell}
&=
\left(
\mathrm{LIF}\!\left[
\begin{tikzpicture}[baseline=-0.5ex,scale=0.92,line cap=round,line join=round]

  \def\xL{-0.38}
  \def\xR{ 0.38}

  \draw[thick] (0,0) circle (1.68);
  \draw[thick] (-0.90, 1.42)--(\xL,1.25);
  \draw[thick] ( 0.90, 1.42)--(\xR,1.25);
  \draw[thick] (-0.90,-1.42)--(\xL,-1.10);
  \draw[thick] ( 0.90,-1.42)--(\xR,-1.10);

  \draw[thick] (\xL,1.25)--(\xL,0.35);
  \draw[thick] (\xR,1.25)--(\xR,0.35);
  \draw[thick] (\xL,1.25)--(\xR,1.25);
  \draw[thick] (\xL,0.80)--(\xR,0.80);
  \draw[thick] (\xL,0.35)--(\xR,0.35);

  \draw[thick] (\xL,0.35)--(\xL,0.12);
  \draw[thick] (\xR,0.35)--(\xR,0.12);

  \node at (0,-0.18) {$\vdots$};

  \draw[thick] (\xL,-0.55)--(\xL,-0.78);
  \draw[thick] (\xR,-0.55)--(\xR,-0.78);

  \draw[thick] (\xL,-0.78)--(\xL,-1.10);
  \draw[thick] (\xR,-0.78)--(\xR,-1.10);
  \draw[thick] (\xL,-0.78)--(\xR,-0.78);
  \draw[thick] (\xL,-1.10)--(\xR,-1.10);

  \draw[thick,dash pattern=on 5pt off 3pt] (-0.95,1.03)--(0.95,1.03);
  \draw[thick,dash pattern=on 5pt off 3pt] (-0.95,0.57)--(0.95,0.57);
  \draw[thick,dash pattern=on 5pt off 3pt] (-0.95,-0.94)--(0.95,-0.94);

\end{tikzpicture}
\right]
\right)^{(1)}_{[\CO^k]_{r,\ell}}
\end{aligned}
\right]
\right|_{\log^{k-2}\ell}.
\label{diaggg50}
\end{equation}

\section{Conclusion}
\label{conclusion}

In this paper we studied four-point functions of identical scalar operators in CFTs with weakly coupled AdS duals in large-$N$ expansion. We discussed how higher-trace operators are required by crossing symmetry from the infinite spin sums of lower-trace data. We also discussed how the different terms in the conformal block expansion can be organized diagrammatically in a way that keeps track of their associated OPE data. This diagrammatic representation also makes the higher-trace appearance and the discussion on infinite spin-sums more clear. 

We now summarize the main results and discuss some directions for future works.

\subsection{Summary of results}

\noindent\textbf{1.}
In Section~\ref{sec1}, we extended the bubble-diagram analysis of \cite{BubblesAdS} from the special case of $\Delta_\CO=2$ to generic external dimension $\Delta_\CO$. The part of the double-trace data fixed by tree-level input can be encoded in an effective double-trace exchange,
\begin{equation}
\widehat{\gamma}_{0,\ell} = \sum_{n,m\geq0} \frac{ f_{nm}}{J^{2\Delta_\CO+2n+2m+\epsilon_n}},
\end{equation}
for the leading twist. We computed the first few coefficients in this expansion, \eqref{f00i}--\eqref{f01i}. Moreover, using the large-spin result of \cite{Kaviraj:2015cxa}, we found the coefficient multiplying the leading large-spin behavior of the full double-trace tower. Namely,
\begin{equation}
    \gamma^{(2)}_{n,\ell} = \frac{\zeta_n}{J^{2\Delta_\CO}} + \mathcal{O}\!\left(\frac{1}{J^{2\Delta_\CO+2}}\right), \qquad \gamma^{(3)}_{n,\ell}\Big|_{\log J} = -\frac{\zeta_n}{J^{2\Delta_\CO}} + \mathcal{O}\!\left(\frac{1}{J^{2\Delta_\CO+2}}\right),
\end{equation}
with the coefficient $\zeta_n$ given in closed-form in~\eqref{zetan}.\newline

\noindent\textbf{2.}
In Section~\ref{sec3}, we showed how higher-trace operators appear by the necessity of crossing symmetry once infinite spin sums are treated carefully. In the $\phi^4$ theory, provided that no mixing between double-trace family and quadruple-trace family occurs, the crossed image of double-trace spin sums forces the appearance of quadruple-trace operators at order $N^{-8}$. For the leading-twist quadruple-trace family, we computed the OPE coefficient $b^{(0)}_{0,\ell}$. For the special case of $\Delta_\CO=\frac52$ we obtained the full spin dependence in closed form.
\begin{equation}
    b^{(0)}_{0,\ell} = \frac{ \sqrt{\pi}\,2^{5-2\ell}(2\ell+9)^2\bigl(\ell(\ell+9)+23\bigr)^2\Gamma(\ell+1) }{ 225 (\ell+5)(\ell+6)(\ell+7)(\ell+8) \Gamma\!\left(\ell+\frac{9}{2}\right) } .
\end{equation}
For generic non-integer $\Delta_\CO$, we found
\begin{equation}
    b^{(0)}_{0,\ell} =\frac{ \sqrt{\pi}\, 2^{-4\Delta_\CO-2\ell+1} \Gamma(2\Delta_\CO)^2 }{ \ell^{3/2}\left[(1-\Delta_\CO)_{2\Delta_\CO}\right]^2} \left[ 1+\mathcal{O}\!\left(\frac{1}{\ell}\right) \right].
\end{equation}
We also computed the corresponding leading-twist anomalous dimension and we obtained
\begin{equation}
    \delta^{(1)}_{0,\ell}=2 .
\end{equation}

\noindent\textbf{3.}
We also studied theories with cubic interactions. In this case, under the no-mixing assumption stated in Section~\ref{sec3}, $k$-trace operators can appear at order $N^{-2k}$. We found the leading large-spin behavior of the leading-twist $k$-trace OPE coefficient,
\begin{equation}
\mathfrak a^{(0)}_{k,0,\ell\gg1}=\frac{(-1)^k\,2^{1-k\Delta_\CO-2\ell}\sqrt{\pi}\,(\tilde\gamma_0)^k}{(k-2)!\,\left[(1-\Delta_\CO)_{\frac{k\Delta_\CO}{2}}\right]^2}\,\frac{\log^{k-2}\ell}{\ell^{3/2}}+\cdots .
\end{equation}\newline

\noindent\textbf{4.}
In Section~\ref{sec4}, we introduced a diagrammatic representation for the conformal-block expansion. We took diagrams inside parentheses to denote products of OPE data, and diagrams enclosed in a circle to denote the corresponding contributions to the correlator. This notation gives a refinement of crossing symmetry at two different levels.

The first level is the crossing-closed orbit of a topology. For a crossing orbit $[\Gamma]$, we wrote
\begin{equation*}
    \sum_{\Gamma'\in[\Gamma]}\sum_{c\in \mathcal C_{\rm v}(\Gamma')}\frac{\mathfrak s_{\rm v}(\Gamma')}{2}\,\circdiag{\Gamma'_c}=\sum_{\Gamma'\in[\Gamma]}\sum_{c\in \mathcal C_{\rm h}(\Gamma')}\frac{\mathfrak s_{\rm h}(\Gamma')}{2}\, \circdiag{\Gamma'_c}\, .
    \tag{\ref{diag35}}
\end{equation*}
This expresses crossing not only for the full correlator, but also for the subset of terms associated with a crossing-closed diagrammatic topology.

The second level is an auxiliary fixed-channel relation written in terms of dot-cuts. Since dot-cuts keep track of the pairing across the cut, they allow one to resolve a fixed channel diagram $\Gamma'$ before recombining the result into dash-cuts. At this auxiliary level we wrote
\begin{equation*}
    \sum_{\dot c\in \dot{\mathcal C}_{\rm v}(\Gamma')} \circdiag{\Gamma'_{\dot c}} = \sum_{\dot c\in \dot{\mathcal C}_{\rm h}(\Gamma')} \circdiag{\Gamma'_{\dot c}}\, .
    \tag{\ref{diag39}}
\end{equation*}

Although this relation is auxiliary, it is still useful. In many examples, dot-cut terms can be recombined into dash-cut terms, up to additional external horizontal cuts. The latter have vanishing crossed-channel double-discontinuity and therefore do not contribute to the Lorentzian inversion formula. In this way, the dot-cut relation refines the use of the inversion formula: it gives a diagrammatic way to identify which crossed-channel terms should be probed in order to extract a given piece of OPE data.\newline

\noindent\textbf{5.}
Using \eqref{diag39} for scalar exchange, we found that the asymptotic and the finite-spin part of the tree-level anomalous dimension have different diagrammatic origins. For $\Delta_\CO=2$:
\begin{equation}
\frac{1}{a^{(0)}_{n,\ell}}
\left(
\begin{tikzpicture}[
    baseline=-0.5ex,
    scale=0.72,
    line cap=round,
    line join=round
]
  \draw[thick] (0,0.32)--(0,-0.32);

  \draw[thick] (-1.00, 0.62)--(0, 0.32);
  \draw[thick] ( 0, 0.32)--(1.00, 0.62);
  \draw[thick] (-1.00,-0.62)--(0,-0.32);
  \draw[thick] ( 0,-0.32)--(1.00,-0.62);

  \draw[thick,dash pattern=on 5pt off 3pt] (-0.62,-0.74)--(-0.62,0.74);
\end{tikzpicture}
\right)^{(0)}_{n,\ell}=
\gamma^{(1)(2)}_{n,\ell}=-\frac{2a_{2,0}}{(\ell+1)(\ell+2n+2)},
\end{equation}

\begin{equation}
\frac{1}{a^{(0)}_{n,\ell}}
\left(
\begin{tikzpicture}[
    baseline=-0.5ex,
    scale=0.72,
    line cap=round,
    line join=round
]
  \coordinate (L) at (-0.35,0);
  \coordinate (R) at ( 0.35,0);

  \draw[thick] (L)--(R);

  \draw[thick] (-1.00, 0.58)--(L);
  \draw[thick] (-1.00,-0.58)--(L);
  \draw[thick] (R)--(1.00, 0.58);
  \draw[thick] (R)--(1.00,-0.58);

  \draw[thick,dash pattern=on 5pt off 3pt] (-0.65,-0.72)--(-0.65,0.72);
\end{tikzpicture}
\right)^{(0)}_{n,\ell}=
\gamma^{(1)(3)}_{n,\ell}=\delta_{\ell,0}\,\frac{1}{2}\,\frac{(n+1)}{(2n+1)(2n+3)}\,a_{2,0}\, .
\end{equation}
As a check of this assignment, we glued the tree-level building blocks in scalar $\phi^3+\phi^4$ theory and computed the corresponding one-loop Mellin amplitudes. We saw that their flat-space limits reproduce the expected massless flat-space amplitudes.\newline

\noindent\textbf{6.}
Finally, we gave a diagrammatic interpretation of the higher-trace analysis. In the $\phi^4$ theory, the diagrammatic representation makes manifest that the leading quadruple-trace OPE coefficient is associated with the infinite double-trace spin sum appearing in~\eqref{dtlog}. This is encoded directly in the diagram~\eqref{b0lif}. In this sense, the diagrammatic representation helps identify which infinite spin sums can generate higher-trace data, and which crossed cuts should be used to extract that data through the Lorentzian inversion formula. Similarly, in the $\phi^3$ theory, the diagrammatic relation~\eqref{diaggg50} makes clear how the higher-trace OPE data are extracted from the crossed-channel terms in~\eqref{general}.

\subsection{Future directions}

\noindent\textbf{1.}
It is important to clarify the bulk interpretation of the cut diagrams introduced here. In \cite{BubblesAdS}, the maximal cuts of bubble diagrams were shown to reduce, in the flat-space limit, to ordinary flat-space unitarity cuts. A first step would be to understand whether an analogous flat-space limit interpretation exists for the more general non-maximal cuts appearing in the present diagrammatic representation. Furthermore, it would be interesting to clarify whether these more general cut diagrams admit a direct interpretation as AdS unitarity cuts, for example along the lines of the construction of~\cite{Meltzer:2019nbs}.\newline

\noindent\textbf{2.}
In the higher-trace analysis we assumed a no-mixing condition between different operator families. It would be important to relax this assumption. Diagrammatically, one expects that mixing should require the inclusion of disconnected contributions. This can also modify the order in the large-$N$ expansion at which some higher-trace data first appear. Understanding this mixing systematically makes the higher-trace analysis more complete.\newline

\noindent\textbf{3.}
It would also be useful to compute more higher-trace data explicitly, including subleading twists, and subleading orders in the large-$N$ expansion. Moreover, we should attempt to resum the large spin expansion of higher-trace data to see whether we can find more closed form expressions for generic $\Delta_\CO$. \newline

\noindent\textbf{4.}
Another important direction is to generalize the product rule in the diagrammatic representation to higher-particle cuts. In this work we formulated the gluing rule for internal two-particle cuts, where two subdiagrams are glued along the same double-trace state by
\begin{equation}
    X\otimes_2 Y=\frac{XY}{a^{(0)}_{n,\ell}}\, .
\end{equation}
For cuts through three, four, or more propagators, the exchanged states are higher-trace families and can carry additional degeneracy labels. A systematic extension should determine the appropriate gluing rule for the higher-trace case. Such a rule would make the diagrammatic organization of higher-trace contributions more complete.

\section*{Acknowledgements}

I am especially grateful to Agnese Bissi for many useful discussions and her detailed comments on the draft. I also thank Ahmadullah Zahed for helpful comments on the draft.

\appendix

\section{Spin Sums}\label{appxA}

In this appendix, we introduce a set of functions that capture the spin sums appearing in the conformal block expansion. We define
\begin{equation} \label{A1}
H_n^{(m)}(z,\bar{z})=\sum_{\ell} a_{n,\ell}^{(0)}\,\frac{(z\bar{z})^{\tau_n/2}}{J^{2m}} \,\tilde{g}_{n,\ell}(z,\bar{z}),
\end{equation}
which represents the general structure of spin sums that appear in the conformal block expansion of double-trace operators.  In the literature, these objects are often referred to as twist conformal blocks \cite{Alday:2016njk,Alday:2016jfr,Aharony:2016dwx,Alday:2017gde}.

Inserting the explicit form of the four-dimensional conformal blocks \eqref{kblock} into \eqref{A1}, we obtain
\begin{equation}\label{decH}
H_n^{(m)}(z,\bar{z})= \frac{1}{z-\bar{z}}\left( \bar{z}^{\tau_n/2} F_{\frac{\tau_n-2}{2}}(\bar{z})\,\tilde{H}_n^{(m)}(z) - z^{\tau_n/2} F_{\frac{\tau_n-2}{2}}(z)\,\tilde{H}_n^{(m)}(\bar{z}) \right),
\end{equation}
where
\begin{equation}\label{A2}
\tilde{H}_n^{(m)}(z)=\sum_{\ell} \frac{a_{n,\ell}^{(0)}}{J^{2m}}\, z^{\tau_n/2+\ell+1}\, F_{\frac{\tau_n}{2}+\ell}(z), \qquad \tau_n=2\Delta_\CO+2n
\end{equation}

Our goal is to determine these functions in the limit $\bar z \to 1$. The functions $\tilde{H}_n^{(m)}(z)$ satisfy a recursive differential equation of the form \cite{Alday:2017gde}
\begin{equation}
D_{4d}\, \tilde{H}_n^{(m+1)}(z)= \tilde{H}_n^{(m)}(z),  \qquad  D_{4d}=z\,D\,z^{-1},
\end{equation}
where the differential operator $D$ is given by
\begin{equation}
D=(1-z)z^2 \partial_z^2 - z^2 \partial_z.
\end{equation}
A similar relation holds for $\tilde{H}_n^{(m)}(\bar{z})$.

Given $\tilde{H}_n^{(0)}(\bar{z})$, the recursive relation allows us to determine $\tilde{H}_n^{(m)}(\bar{z})$ for all $m$. We now review the steps of \cite{Alday:2017gde} to extract $\tilde{H}_n^{(0)}(\bar{z})$.

$\tilde{H}_n^{(0)}(\bar{z})$ appears in the conformal block expansion of the generalized free field correlator,
\begin{equation}
\mathcal{G}^{(0)}(z,\bar{z})=1+\frac{1}{z-\bar{z}}\Big( \sum_{n,\ell}a^{(0)}_{n,\ell}\, z\bar{z} \, k_{\frac{\tau_n}{2}+\ell}(z)k_{\frac{\tau_n}{2}-1}(\bar{z})-\sum_n z \, k_{\frac{\tau_n}{2}-1}(z) \tilde{H}_n^{(0)}(\bar{z})\Big).
\end{equation}

To isolate $\tilde{H}_n^{(0)}(\bar{z})$, we use the orthogonality relation
\begin{equation}
\oint_{z=0} \frac{dz}{2\pi i} \frac{1}{z^2} k_{\frac{\tau_n}{2}-1}(z)k_{2-\frac{\tau_m}{2}}(z)=\delta_{n,m}.
\end{equation}
This yields
\begin{equation}
\tilde{H}_n^{(0)}(\bar{z})= B_n^{(0)}(\bar{z})+P_n^{(0)}(\bar{z}),
\end{equation}
where
\begin{equation}
B_n^{(0)}(\bar{z})= \sum_{m,\ell} \delta_{m+\ell,n-1}\, a^{(0)}_{m,\ell}\, \bar{z}\, k_{\frac{\tau_m}{2}-1}(\bar{z}),
\end{equation}
and
\begin{equation}\label{pn}
P_n^{(0)}(\bar{z})= -\oint_{z=0} \frac{dz}{2\pi i}\frac{1}{z^3} k_{2-\frac{\tau_n}{2}}(z) (z-\bar{z})\Big(\mathcal{G}^{(0)}(z,\bar{z})-1\Big).
\end{equation}

Note that
\begin{equation}
\bar{D}_{4d}\, \bar{z}\, k_\alpha (\bar{z})=\alpha(\alpha-1)\, \bar{z}\, k_\alpha (\bar{z}).
\end{equation}
This implies that the functions $B_n^{(m)}(\bar{z})$ can be constructed iteratively such that
\begin{equation}
\bar{D}_{4d}\, B_n^{(m+1)}(\bar{z})=B_n^{(m)}(\bar{z}).
\end{equation}
So we can also define $P_n^{(m)}(\bar{z})$ to satisfy
\begin{equation} \label{Dpn}
\bar{D}_{4d}\, P_n^{(m+1)}(\bar{z})=P_n^{(m)}(\bar{z}).
\end{equation}

The functions $B_n^{(m)}(\bar{z})$ exhibit the same behavior as a single conformal block in the limit $\bar{z} \to 1$. In contrast, $P_n^{(m)}(\bar{z})$ can display new behavior, arising from the infinite sum over spins.

From the definition of $P_n^{(0)}$, we find that near $\bar{z}=1$
\begin{equation}\label{414}
P_n^{(0)}(\bar{z})=\frac{Y_n}{(1-\bar{z})^{\Delta_\mathcal{O}}} + \mathcal{O}\big((1-\bar{z})^{-\Delta_\mathcal{O}+1}\big).
\end{equation}
where
\begin{equation}\label{yn}
Y_n= \frac{4^{-n} (\Delta_\mathcal{O}-1)_n (2 \Delta_\mathcal{O}-3)_n}{(1)_n \left(\Delta_\mathcal{O}-\frac{3}{2}\right)_n}.
\end{equation}

Using the recursion relation~\eqref{Dpn}, we can determine the behavior of $P_n^{(m)}(\bar{z})$ near $\bar{z}=1$. For $\Delta_\mathcal{O}-m \notin \mathbb{Z}_{\leq 0}$, we find
\begin{equation}\label{ppn}
P_n^{(m)}(\bar{z})= \frac{Y_n}{\big((1-\Delta_\mathcal{O})_m\big)^2}(1-\bar{z})^{-\Delta_\mathcal{O}+m}+\mathcal{O}\!\left((1-\bar{z})^{-\Delta_\mathcal{O}+m+1}\right).
\end{equation}

In the case $\Delta_\mathcal{O}-m \in \mathbb{Z}_{\leq 0}$, we parametrize $m=\Delta_\mathcal{O}+p$ with $p \in \mathbb{Z}_{\geq 0}$. In such case, near $\bar{z}=1$ we find
\begin{equation}
P_n^{(m=\Delta_\mathcal{O}+p)}(\bar{z})= \frac{1}{2(p!)^2}\frac{Y_n}{\Gamma^2(\Delta_\mathcal{O})} (1-\bar{z})^p \log^2(1-\bar{z}) +\mathcal{O}\!\big((1-\bar{z})^p \log(1-\bar{z})\big).
\end{equation}

Finally, we note that in some cases we encounter a more general class of spin sums of the form
\begin{equation}
\tilde{H}_n^{(m)(q)} (z) = \sum_{\ell} a_{n,\ell}\frac{\log^q J}{J^{2m}}\, z^{\tau_n/2+\ell+1}\, F_{\frac{\tau_n}{2}+\ell}(z),
\end{equation}
such that the previously defined functions correspond to the case $q=0$, i.e.
\begin{equation}
\tilde{H}_n^{(m)} (z)= \tilde{H}_n^{(m)(0)} (z).
\end{equation}

For $q=1$, these functions can be written as
\begin{align}
\tilde{H}_n^{(m)(1)} (z)&= -\frac{1}{2}\,\frac{\partial}{\partial a}\sum_{\ell} a_{n,\ell}\frac{1}{J^{2(m+a)}}\, z^{\tau_n/2+\ell+1}\, F_{\frac{\tau_n}{2}+\ell}(z)\Big|_{a=0} \\&= -\frac{1}{2}\,\frac{\partial}{\partial a}\, \tilde{H}_n^{(m+a)} (z)\Big|_{a=0} \\&= -\frac{1}{2}\,\frac{\partial}{\partial a}\Big( B_n^{(m+a)}(z)+P_n^{(m+a)}(z)\Big)\Big|_{a=0}.
\end{align}

This allows us to define the corresponding decompositions $B_n^{(m)(1)}$ and $P_n^{(m)(1)}$. In the case $\Delta_\mathcal{O}-m \notin \mathbb{Z}_{\leq 0}$, we find for the leading behavior near $\bar{z}=1$
\begin{equation}\label{wlog}
\begin{aligned}
P_n^{(m)(1)}(\bar{z})
&= -\frac{Y_n}{2\big((1-\Delta_\mathcal{O})_m\big)^2} (1-\bar{z})^{-\Delta_\mathcal{O}+m} \left[\log(1-\bar{z})-2\,\psi(1-\Delta_\mathcal{O}+m)\right] \\
&\quad +\mathcal{O}\!\left((1-\bar{z})^{-\Delta_\mathcal{O}+m+1}\log(1-\bar{z})\right).
\end{aligned}
\end{equation}
where $\psi(x)$ denotes the digamma function.

\section{Mellin space}\label{app2}

\subsection{Mellin amplitude}

An alternative way to represent correlation functions in CFT is provided by Mellin space, which naturally provides an interpretation in terms of AdS amplitudes. \cite{Mack:2009mi,Mack:2009gy,Penedones:2010ue,Fitzpatrick:2011ia,Paulos:2011ie,Nandan:2011wc}. Mellin space is also used in Polyakov-Mellin bootstrap methods and in dispersion relations for CFT correlators~\cite{Gopakumar:2016wkt,Gopakumar:2016cpb,Gopakumar:2018xqi,Ghosh:2019lsx,Ferrero:2019luz,Penedones:2019tng,Carmi:2020ekr,Caron-Huot:2020adz,Gopakumar:2021dvg}.

For a given four-point function, a Mellin amplitude $M(s,t)$ is defined by the relation
\begin{equation}
\mathcal{G}(u,v) = \int_{-i\infty}^{i\infty}\frac{ds\,dt}{(4\pi i)^2}\,u^{\frac{t}{2}}\,v^{\frac{\hat u-2\Delta_\CO}{2}}\,\Gamma^2\!\left(\frac{2\Delta_\CO-t}{2}\right)\Gamma^2\!\left(\frac{2\Delta_\CO-s}{2}\right)\Gamma^2\!\left(\frac{2\Delta_\CO-\hat u}{2}\right)M(s,t),
\end{equation}
where $s,t,\hat u$ are Mandelstam-like variables satisfying
\begin{equation}
s+t+\hat u=4\Delta_\CO .
\end{equation}
Crossing symmetry in Mellin space reads
\begin{equation}
M(s,t)=M(s,\hat u)=M(t,s).
\end{equation}

Similarly to the correlator $\mathcal{G}(u,v)$, the Mellin amplitude $M(s,t)$ admits a large-$N$ expansion,
\begin{equation}
M(s,t)=\frac{M_{\text{tree}}(s,t)}{N^2}+\frac{M_{1\text{-loop}}(s,t)}{N^4}+\cdots .
\end{equation}

Analogously to the conformal block decomposition, the polar part of the Mellin amplitude in the $t$-channel can be written as
\begin{equation}
M(s,t) = \sum_p a_{\Delta_p,\ell_p} \sum_{m=0}^{\infty} \frac{\mathcal{Q}_{\ell_p,m}(s,\tau_p)} {t-(\tau_p+2m)} .
\end{equation}
The sum over $p$ runs over the primary operators $\CO_p$, with twist $\tau_p$ and spin $\ell_p$, in the $\CO\times\CO$ OPE. The sum over $m$ accounts for the descendants, and the functions $\mathcal{Q}_{\ell_p,m}(s,\tau_p)$ are Mack polynomials \cite{Mack:2009gy} and they can be written as \cite{Costa:2012cb}
\begin{equation}
\CQ_{\ell_p,m}(s,\tau_p) = \frac{-2\,\Gamma(\Delta_p+\ell_p)(\Delta_p-1)_{\ell_p}}{4^{\ell_p}\,m!(\Delta_p-1)_m\,\Gamma^4\left(\frac{\Delta_p+\ell_p}{2}\right)\Gamma^2\!\left(\Delta_\CO-m-\frac{\Delta_p-\ell_p}{2}\right)}\,Q_{\ell_p,m}(s,\tau_p),
\end{equation}
where $Q_{\ell_p,m}(s,\tau_p)$ is a polynomial of degree $\ell_p$ in $s$, with $Q_{0,m}(s,\tau_p)=1.$

Specializing to the case where the exchanged operators $\CO_p$ are double-trace operators, we have
\begin{equation}
\CQ_{\ell,m}\!\left(s,2\Delta_\CO+2n+\frac{\gamma^{(1)}_{n,\ell}}{N^2}+\frac{\gamma^{(2)}_{n,\ell}}{N^4}+\cdots\right)\propto\frac{\left(\gamma^{(1)}_{n,\ell}\right)^2}{N^4}\,Q_{\ell,m}(s,2\Delta_\CO+2n)+\mathcal O(N^{-6}) .
\end{equation}

Writing
\begin{equation}
    \gamma^{(1)}_{n,\ell}=\sum_i \gamma^{(1)(i)}_{n,\ell},
\end{equation}
we therefore have
\begin{equation}
    \left(\gamma^{(1)}_{n,\ell}\right)^2 = \sum_{i,j}\gamma^{(1)(i)}_{n,\ell}\,\gamma^{(1)(j)}_{n,\ell}.
\end{equation}
Here the superscripts $(i)$ and $(j)$ label the different diagrams contributing to $\gamma^{(1)}_{n,\ell}$, as introduced in \eqref{diag25}, \eqref{diag26}, and \eqref{diag27}. In the application considered in Section~\ref{lastsub}, the relevant products are nonzero only for $\ell=0$. Therefore we find\footnote{We are excluding the case of $(i)(j)=(2)(2)$.}
\begin{equation}\label{mmamp}
[M_{1\text{-loop}}(s,t)]_{(i)(j)} =\sum_{m=0}^{\infty}\frac{[R_m^{(1)}]_{(i)(j)}}{t-(2\Delta_\CO+2m)} ,
\end{equation}
with
\begin{equation}\label{rramp}
[R_m^{(1)}]_{(i)(j)}=\sum_{n=0}^m-\frac{4(2 n+1)(2 n+3)\Gamma(m+1)^2}{\Gamma(m-n+1)\Gamma(m+n+3)}\,\gamma^{(1)(i)}_{n,0}\,\gamma^{(1)(j)}_{n,0},
\end{equation}
for  $\Delta_\CO=2$. Different choices of the pair $(i)(j)$ correspond to the different diagrams
considered in Section~\ref{lastsub}.

\subsection{Flat-space limit}

The flat-space scattering amplitude can be reproduced holographically by taking the AdS radius to infinity. Such a relation is best realized at the level of Mellin amplitudes \cite{Penedones:2010ue, Fitzpatrick:2011ia}. Namely\footnote{with our conventions $s_{12}=t, s_{13}=s, s_{14}=\hat u$.}

\begin{subequations}
\begingroup
\setlength{\abovedisplayskip}{3pt}
\setlength{\belowdisplayskip}{3pt}

\begin{equation}
\resizebox{0.96\textwidth}{!}{$
\displaystyle
M(s_{ij})\approx \frac{R_{\rm AdS}^{-d+3}}{\CN \Gamma\mleft(\frac{1}{2}\sum_i \Delta_i-\frac{d}{2}\mright)}\int_{0}^\infty \!\!\!d\beta\lsp  \beta^{\frac{1}{2}\sum_i \Delta_i-\frac{d}{2}-1}e^{-\beta} A\left(S_{ij}=\frac{2\beta}{R^2_{\rm AdS}}s_{ij}\right), \,  s_{ij}\gg 1,
$}
\end{equation}

\vspace{-0.8em}

\begin{equation}\label{flatlimit}
\resizebox{0.96\textwidth}{!}{$
\displaystyle
A(S_{ij})=\frac{\CN \Gamma\mleft(\frac{1}{2}\sum_i \Delta_i-\frac{d}{2}\mright)}{R_{\rm AdS}^{-d+3}} \!\!\lim_{R_{\rm AdS}\to \infty}\int_{-i \infty}^{+i \infty}\frac{d\alpha}{2\pi i }\alpha^{\frac{d}{2}-\frac{1}{2}\sum_i \Delta_i}e^\alpha  M\left(s_{ij}=\frac{R_{\rm AdS}^2}{2\alpha}S_{ij}\right),
$}
\end{equation}

\endgroup
\end{subequations}

where $R_{\text{AdS}}$ is the radius of AdS$_{d+1}$, and $\mathcal N$ is a normalization constant which, in our normalization, we take $\mathcal N=8\pi^2$.

The quantity $A(S_{ij})$ is the flat-space scattering amplitude.

In the body of the paper we used this relation to check some of the diagrammatic interpretations.

\addtocontents{toc}{\protect\vskip1.5em}

\Bibliography[refs.bib]

\providecommand{\href}[2]{#2}\begingroup\raggedright\begin{thebibliography}{10}

\bibitem{Maldacena:1997re}
J.M.~Maldacena, \emph{{The Large $N$ limit of superconformal field theories and supergravity}}, \href{https://doi.org/10.4310/ATMP.1998.v2.n2.a1}{\emph{Adv. Theor. Math. Phys.} {\bfseries 2} (1998) 231} [\href{https://arxiv.org/abs/hep-th/9711200}{{\ttfamily hep-th/9711200}}].

\bibitem{Witten:1998qj}
E.~Witten, \emph{{Anti de Sitter space and holography}}, \href{https://doi.org/10.4310/ATMP.1998.v2.n2.a2}{\emph{Adv. Theor. Math. Phys.} {\bfseries 2} (1998) 253} [\href{https://arxiv.org/abs/hep-th/9802150}{{\ttfamily hep-th/9802150}}].

\bibitem{Gubser:1998bc}
S.S.~Gubser, I.R.~Klebanov and A.M.~Polyakov, \emph{{Gauge theory correlators from noncritical string theory}}, \href{https://doi.org/10.1016/S0370-2693(98)00377-3}{\emph{Phys. Lett. B} {\bfseries 428} (1998) 105} [\href{https://arxiv.org/abs/hep-th/9802109}{{\ttfamily hep-th/9802109}}].

\bibitem{Heemskerk:2009pn}
I.~Heemskerk, J.~Penedones, J.~Polchinski and J.~Sully, \emph{{Holography from Conformal Field Theory}}, \href{https://doi.org/10.1088/1126-6708/2009/10/079}{\emph{JHEP} {\bfseries 10} (2009) 079} [\href{https://arxiv.org/abs/0907.0151}{{\ttfamily 0907.0151}}].

\bibitem{Penedones:2010ue}
J.~Penedones, \emph{{Writing CFT correlation functions as AdS scattering amplitudes}}, \href{https://doi.org/10.1007/JHEP03(2011)025}{\emph{JHEP} {\bfseries 03} (2011) 025} [\href{https://arxiv.org/abs/1011.1485}{{\ttfamily 1011.1485}}].

\bibitem{Fitzpatrick:2011ia}
A.L.~Fitzpatrick, J.~Kaplan, J.~Penedones, S.~Raju and B.C.~van Rees, \emph{{A Natural Language for AdS/CFT Correlators}}, \href{https://doi.org/10.1007/JHEP11(2011)095}{\emph{JHEP} {\bfseries 11} (2011) 095} [\href{https://arxiv.org/abs/1107.1499}{{\ttfamily 1107.1499}}].

\bibitem{Fitzpatrick:2011hu}
A.L.~Fitzpatrick and J.~Kaplan, \emph{{Analyticity and the Holographic S-Matrix}}, \href{https://doi.org/10.1007/JHEP10(2012)127}{\emph{JHEP} {\bfseries 10} (2012) 127} [\href{https://arxiv.org/abs/1111.6972}{{\ttfamily 1111.6972}}].

\bibitem{Fitzpatrick:2011dm}
A.L.~Fitzpatrick and J.~Kaplan, \emph{{Unitarity and the Holographic S-Matrix}}, \href{https://doi.org/10.1007/JHEP10(2012)032}{\emph{JHEP} {\bfseries 10} (2012) 032} [\href{https://arxiv.org/abs/1112.4845}{{\ttfamily 1112.4845}}].

\bibitem{El-Showk:2011yvt}
S.~El-Showk and K.~Papadodimas, \emph{{Emergent Spacetime and Holographic CFTs}}, \href{https://doi.org/10.1007/JHEP10(2012)106}{\emph{JHEP} {\bfseries 10} (2012) 106} [\href{https://arxiv.org/abs/1101.4163}{{\ttfamily 1101.4163}}].

\bibitem{Fitzpatrick:2012yx}
A.L.~Fitzpatrick, J.~Kaplan, D.~Poland and D.~Simmons-Duffin, \emph{{The Analytic Bootstrap and AdS Superhorizon Locality}}, \href{https://doi.org/10.1007/JHEP12(2013)004}{\emph{JHEP} {\bfseries 12} (2013) 004} [\href{https://arxiv.org/abs/1212.3616}{{\ttfamily 1212.3616}}].

\bibitem{Komargodski:2012ek}
Z.~Komargodski and A.~Zhiboedov, \emph{{Convexity and Liberation at Large Spin}}, \href{https://doi.org/10.1007/JHEP11(2013)140}{\emph{JHEP} {\bfseries 11} (2013) 140} [\href{https://arxiv.org/abs/1212.4103}{{\ttfamily 1212.4103}}].

\bibitem{Fitzpatrick:2014vua}
A.L.~Fitzpatrick, J.~Kaplan and M.T.~Walters, \emph{{Universality of Long-Distance AdS Physics from the CFT Bootstrap}}, \href{https://doi.org/10.1007/JHEP08(2014)145}{\emph{JHEP} {\bfseries 08} (2014) 145} [\href{https://arxiv.org/abs/1403.6829}{{\ttfamily 1403.6829}}].

\bibitem{Alday:2014tsa}
L.F.~Alday, A.~Bissi and T.~Lukowski, \emph{{Lessons from crossing symmetry at large N}}, \href{https://doi.org/10.1007/JHEP06(2015)074}{\emph{JHEP} {\bfseries 06} (2015) 074} [\href{https://arxiv.org/abs/1410.4717}{{\ttfamily 1410.4717}}].

\bibitem{Kaviraj:2015cxa}
A.~Kaviraj, K.~Sen and A.~Sinha, \emph{{Analytic bootstrap at large spin}}, \href{https://doi.org/10.1007/JHEP11(2015)083}{\emph{JHEP} {\bfseries 11} (2015) 083} [\href{https://arxiv.org/abs/1502.01437}{{\ttfamily 1502.01437}}].

\bibitem{Kaviraj:2015xsa}
A.~Kaviraj, K.~Sen and A.~Sinha, \emph{{Universal anomalous dimensions at large spin and large twist}}, \href{https://doi.org/10.1007/JHEP07(2015)026}{\emph{JHEP} {\bfseries 07} (2015) 026} [\href{https://arxiv.org/abs/1504.00772}{{\ttfamily 1504.00772}}].

\bibitem{Alday:2015eya}
L.F.~Alday, A.~Bissi and T.~Lukowski, \emph{{Large spin systematics in CFT}}, \href{https://doi.org/10.1007/JHEP11(2015)101}{\emph{JHEP} {\bfseries 11} (2015) 101} [\href{https://arxiv.org/abs/1502.07707}{{\ttfamily 1502.07707}}].

\bibitem{Alday:2016njk}
L.F.~Alday, \emph{{Large Spin Perturbation Theory for Conformal Field Theories}}, \href{https://doi.org/10.1103/PhysRevLett.119.111601}{\emph{Phys. Rev. Lett.} {\bfseries 119} (2017) 111601} [\href{https://arxiv.org/abs/1611.01500}{{\ttfamily 1611.01500}}].

\bibitem{Aharony:2016dwx}
O.~Aharony, L.F.~Alday, A.~Bissi and E.~Perlmutter, \emph{{Loops in AdS from Conformal Field Theory}}, \href{https://doi.org/10.1007/JHEP07(2017)036}{\emph{JHEP} {\bfseries 07} (2017) 036} [\href{https://arxiv.org/abs/1612.03891}{{\ttfamily 1612.03891}}].

\bibitem{Alday:2017gde}
L.F.~Alday, A.~Bissi and E.~Perlmutter, \emph{{Holographic Reconstruction of AdS Exchanges from Crossing Symmetry}}, \href{https://doi.org/10.1007/JHEP08(2017)147}{\emph{JHEP} {\bfseries 08} (2017) 147} [\href{https://arxiv.org/abs/1705.02318}{{\ttfamily 1705.02318}}].

\bibitem{Poland:2018epd}
D.~Poland, S.~Rychkov and A.~Vichi, \emph{{The Conformal Bootstrap: Theory, Numerical Techniques, and Applications}}, \href{https://doi.org/10.1103/RevModPhys.91.015002}{\emph{Rev. Mod. Phys.} {\bfseries 91} (2019) 015002} [\href{https://arxiv.org/abs/1805.04405}{{\ttfamily 1805.04405}}].

\bibitem{Albayrak:2019gnz}
S.~Albayrak, D.~Meltzer and D.~Poland, \emph{{More Analytic Bootstrap: Nonperturbative Effects and Fermions}}, \href{https://doi.org/10.1007/JHEP08(2019)040}{\emph{JHEP} {\bfseries 08} (2019) 040} [\href{https://arxiv.org/abs/1904.00032}{{\ttfamily 1904.00032}}].

\bibitem{Bissi:2022mrs}
A.~Bissi, A.~Sinha and X.~Zhou, \emph{{Selected topics in analytic conformal bootstrap: A guided journey}}, \href{https://doi.org/10.1016/j.physrep.2022.09.004}{\emph{Phys. Rept.} {\bfseries 991} (2022) 1} [\href{https://arxiv.org/abs/2202.08475}{{\ttfamily 2202.08475}}].

\bibitem{Huang:2023oxf}
Z.~Huang, B.~Wang, E.Y.~Yuan and X.~Zhou, \emph{{AdS super gluon scattering up to two loops: a position space approach}}, \href{https://doi.org/10.1007/JHEP07(2023)053}{\emph{JHEP} {\bfseries 07} (2023) 053} [\href{https://arxiv.org/abs/2301.13240}{{\ttfamily 2301.13240}}].

\bibitem{BubblesAdS}
A.~Bissi, G.~Fardelli and M.R.~Khansari, \emph{{Bubbles in AdS}}, \href{https://doi.org/10.1007/JHEP03(2026)246}{\emph{JHEP} {\bfseries 03} (2026) 246} [\href{https://arxiv.org/abs/2509.13036}{{\ttfamily 2509.13036}}].

\bibitem{Bertan:2018afl}
I.~Bertan, I.~Sachs and E.D.~Skvortsov, \emph{{Quantum $\phi^4$ Theory in AdS${}_4$ and its CFT Dual}}, \href{https://doi.org/10.1007/JHEP02(2019)099}{\emph{JHEP} {\bfseries 02} (2019) 099} [\href{https://arxiv.org/abs/1810.00907}{{\ttfamily 1810.00907}}].

\bibitem{Carmi:2026spv}
D.~Carmi, R.~Ciccone and S.~Sukholuski, \emph{{Closing the loop on $\Phi^4$ in {AdS}$_3$}},  \href{https://arxiv.org/abs/2606.06589}{{\ttfamily 2606.06589}}.

\bibitem{Akhmedov:2018lkp}
E.T.~Akhmedov, U.~Moschella and F.K.~Popov, \emph{{Ultraviolet phenomena in AdS self-interacting quantum field theory}}, \href{https://doi.org/10.1007/JHEP03(2018)183}{\emph{JHEP} {\bfseries 03} (2018) 183} [\href{https://arxiv.org/abs/1802.02955}{{\ttfamily 1802.02955}}].

\bibitem{Cacciatori:2024zbe}
S.L.~Cacciatori, H.~Epstein and U.~Moschella, \emph{{Loops in anti de Sitter space}}, \href{https://doi.org/10.1007/JHEP08(2024)109}{\emph{JHEP} {\bfseries 08} (2024) 109} [\href{https://arxiv.org/abs/2403.13142}{{\ttfamily 2403.13142}}].

\bibitem{Carmi:2019ocp}
D.~Carmi, \emph{{Loops in AdS: From the Spectral Representation to Position Space}}, \href{https://doi.org/10.1007/JHEP06(2020)049}{\emph{JHEP} {\bfseries 06} (2020) 049} [\href{https://arxiv.org/abs/1910.14340}{{\ttfamily 1910.14340}}].

\bibitem{Carmi:2021dsn}
D.~Carmi, \emph{{Loops in AdS: from the spectral representation to position space. Part II}}, \href{https://doi.org/10.1007/JHEP07(2021)186}{\emph{JHEP} {\bfseries 07} (2021) 186} [\href{https://arxiv.org/abs/2104.10500}{{\ttfamily 2104.10500}}].

\bibitem{Carmi:2024tzj}
D.~Carmi, \emph{{Loops in AdS: from the spectral representation to position space. Part III}}, \href{https://doi.org/10.1007/JHEP08(2024)193}{\emph{JHEP} {\bfseries 08} (2024) 193} [\href{https://arxiv.org/abs/2402.02481}{{\ttfamily 2402.02481}}].

\bibitem{Bertan:2018khc}
I.~Bertan and I.~Sachs, \emph{{Loops in Anti{\textendash}de Sitter Space}}, \href{https://doi.org/10.1103/PhysRevLett.121.101601}{\emph{Phys. Rev. Lett.} {\bfseries 121} (2018) 101601} [\href{https://arxiv.org/abs/1804.01880}{{\ttfamily 1804.01880}}].

\bibitem{Xiao:2026prw}
W.~Xiao and I.~Sachs, \emph{{The 2-Dimensional Dual of $\phi^4$ in AdS$_3$}},  \href{https://arxiv.org/abs/2602.05750}{{\ttfamily 2602.05750}}.

\bibitem{Albayrak:2020bso}
S.~Albayrak and S.~Kharel, \emph{{Spinning loop amplitudes in anti{\textendash}de Sitter space}}, \href{https://doi.org/10.1103/PhysRevD.103.026004}{\emph{Phys. Rev. D} {\bfseries 103} (2021) 026004} [\href{https://arxiv.org/abs/2006.12540}{{\ttfamily 2006.12540}}].

\bibitem{Caron-Huot:2017vep}
S.~Caron-Huot, \emph{{Analyticity in Spin in Conformal Theories}}, \href{https://doi.org/10.1007/JHEP09(2017)078}{\emph{JHEP} {\bfseries 09} (2017) 078} [\href{https://arxiv.org/abs/1703.00278}{{\ttfamily 1703.00278}}].

\bibitem{Bissi:2020wtv}
A.~Bissi, G.~Fardelli and A.~Georgoudis, \emph{{Towards all loop supergravity amplitudes on AdS5\texttimes{}S5}}, \href{https://doi.org/10.1103/PhysRevD.104.L041901}{\emph{Phys. Rev. D} {\bfseries 104} (2021) L041901} [\href{https://arxiv.org/abs/2002.04604}{{\ttfamily 2002.04604}}].

\bibitem{Bissi:2020woe}
A.~Bissi, G.~Fardelli and A.~Georgoudis, \emph{{All loop structures in supergravity amplitudes on AdS5 \texttimes{} S5 from CFT}}, \href{https://doi.org/10.1088/1751-8121/ac0ebf}{\emph{J. Phys. A} {\bfseries 54} (2021) 324002} [\href{https://arxiv.org/abs/2010.12557}{{\ttfamily 2010.12557}}].

\bibitem{Fardelli:2025eun}
G.~Fardelli, A.L.~Fitzpatrick and W.~Li, \emph{{Towards large-spin effective theory. Part I. Three-particle states in AdS {\ensuremath{\phi}}$^{4}$ theory}}, \href{https://doi.org/10.1007/JHEP06(2026)004}{\emph{JHEP} {\bfseries 06} (2026) 004} [\href{https://arxiv.org/abs/2508.20158}{{\ttfamily 2508.20158}}].

\bibitem{Kravchuk:2024wmv}
P.~Kravchuk and J.A.~Mann, \emph{{AdS N-body problem at large spin}}, \href{https://doi.org/10.1007/JHEP10(2025)004}{\emph{JHEP} {\bfseries 10} (2025) 004} [\href{https://arxiv.org/abs/2412.12328}{{\ttfamily 2412.12328}}].

\bibitem{Dolan:2003hv}
F.A.~Dolan and H.~Osborn, \emph{{Conformal partial waves and the operator product expansion}}, \href{https://doi.org/10.1016/j.nuclphysb.2003.11.016}{\emph{Nucl. Phys. B} {\bfseries 678} (2004) 491} [\href{https://arxiv.org/abs/hep-th/0309180}{{\ttfamily hep-th/0309180}}].

\bibitem{Meltzer:2019nbs}
D.~Meltzer, E.~Perlmutter and A.~Sivaramakrishnan, \emph{{Unitarity Methods in AdS/CFT}}, \href{https://doi.org/10.1007/JHEP03(2020)061}{\emph{JHEP} {\bfseries 03} (2020) 061} [\href{https://arxiv.org/abs/1912.09521}{{\ttfamily 1912.09521}}].

\bibitem{Balasubramanian:2001nh}
V.~Balasubramanian, M.~Berkooz, A.~Naqvi and M.J.~Strassler, \emph{{Giant gravitons in conformal field theory}}, \href{https://doi.org/10.1088/1126-6708/2002/04/034}{\emph{JHEP} {\bfseries 04} (2002) 034} [\href{https://arxiv.org/abs/hep-th/0107119}{{\ttfamily hep-th/0107119}}].

\bibitem{Witten:2001ua}
E.~Witten, \emph{{Multitrace operators, boundary conditions, and AdS / CFT correspondence}},  \href{https://arxiv.org/abs/hep-th/0112258}{{\ttfamily hep-th/0112258}}.

\bibitem{Alday:2016jfr}
L.F.~Alday, \emph{{Solving CFTs with Weakly Broken Higher Spin Symmetry}}, \href{https://doi.org/10.1007/JHEP10(2017)161}{\emph{JHEP} {\bfseries 10} (2017) 161} [\href{https://arxiv.org/abs/1612.00696}{{\ttfamily 1612.00696}}].

\bibitem{Mack:2009mi}
G.~Mack, \emph{{D-independent representation of Conformal Field Theories in D dimensions via transformation to auxiliary Dual Resonance Models. Scalar amplitudes}},  \href{https://arxiv.org/abs/0907.2407}{{\ttfamily 0907.2407}}.

\bibitem{Mack:2009gy}
G.~Mack, \emph{{D-dimensional Conformal Field Theories with anomalous dimensions as Dual Resonance Models}}, {\emph{Bulg. J. Phys.} {\bfseries 36} (2009) 214} [\href{https://arxiv.org/abs/0909.1024}{{\ttfamily 0909.1024}}].

\bibitem{Paulos:2011ie}
M.F.~Paulos, \emph{{Towards Feynman rules for Mellin amplitudes}}, \href{https://doi.org/10.1007/JHEP10(2011)074}{\emph{JHEP} {\bfseries 10} (2011) 074} [\href{https://arxiv.org/abs/1107.1504}{{\ttfamily 1107.1504}}].

\bibitem{Nandan:2011wc}
D.~Nandan, A.~Volovich and C.~Wen, \emph{{On Feynman Rules for Mellin Amplitudes in AdS/CFT}}, \href{https://doi.org/10.1007/JHEP05(2012)129}{\emph{JHEP} {\bfseries 05} (2012) 129} [\href{https://arxiv.org/abs/1112.0305}{{\ttfamily 1112.0305}}].

\bibitem{Gopakumar:2016wkt}
R.~Gopakumar, A.~Kaviraj, K.~Sen and A.~Sinha, \emph{{Conformal Bootstrap in Mellin Space}}, \href{https://doi.org/10.1103/PhysRevLett.118.081601}{\emph{Phys. Rev. Lett.} {\bfseries 118} (2017) 081601} [\href{https://arxiv.org/abs/1609.00572}{{\ttfamily 1609.00572}}].

\bibitem{Gopakumar:2016cpb}
R.~Gopakumar, A.~Kaviraj, K.~Sen and A.~Sinha, \emph{{A Mellin space approach to the conformal bootstrap}}, \href{https://doi.org/10.1007/JHEP05(2017)027}{\emph{JHEP} {\bfseries 05} (2017) 027} [\href{https://arxiv.org/abs/1611.08407}{{\ttfamily 1611.08407}}].

\bibitem{Gopakumar:2018xqi}
R.~Gopakumar and A.~Sinha, \emph{{On the Polyakov-Mellin bootstrap}}, \href{https://doi.org/10.1007/JHEP12(2018)040}{\emph{JHEP} {\bfseries 12} (2018) 040} [\href{https://arxiv.org/abs/1809.10975}{{\ttfamily 1809.10975}}].

\bibitem{Ghosh:2019lsx}
K.~Ghosh, \emph{{Polyakov-Mellin Bootstrap for AdS loops}}, \href{https://doi.org/10.1007/JHEP02(2020)006}{\emph{JHEP} {\bfseries 02} (2020) 006} [\href{https://arxiv.org/abs/1811.00504}{{\ttfamily 1811.00504}}].

\bibitem{Ferrero:2019luz}
P.~Ferrero, K.~Ghosh, A.~Sinha and A.~Zahed, \emph{{Crossing symmetry, transcendentality and the Regge behaviour of 1d CFTs}}, \href{https://doi.org/10.1007/JHEP07(2020)170}{\emph{JHEP} {\bfseries 07} (2020) 170} [\href{https://arxiv.org/abs/1911.12388}{{\ttfamily 1911.12388}}].

\bibitem{Penedones:2019tng}
J.~Penedones, J.A.~Silva and A.~Zhiboedov, \emph{{Nonperturbative Mellin Amplitudes: Existence, Properties, Applications}}, \href{https://doi.org/10.1007/JHEP08(2020)031}{\emph{JHEP} {\bfseries 08} (2020) 031} [\href{https://arxiv.org/abs/1912.11100}{{\ttfamily 1912.11100}}].

\bibitem{Carmi:2020ekr}
D.~Carmi, J.~Penedones, J.A.~Silva and A.~Zhiboedov, \emph{{Applications of dispersive sum rules: $\epsilon$-expansion and holography}}, \href{https://doi.org/10.21468/SciPostPhys.10.6.145}{\emph{SciPost Phys.} {\bfseries 10} (2021) 145} [\href{https://arxiv.org/abs/2009.13506}{{\ttfamily 2009.13506}}].

\bibitem{Caron-Huot:2020adz}
S.~Caron-Huot, D.~Mazac, L.~Rastelli and D.~Simmons-Duffin, \emph{{Dispersive CFT Sum Rules}}, \href{https://doi.org/10.1007/JHEP05(2021)243}{\emph{JHEP} {\bfseries 05} (2021) 243} [\href{https://arxiv.org/abs/2008.04931}{{\ttfamily 2008.04931}}].

\bibitem{Gopakumar:2021dvg}
R.~Gopakumar, A.~Sinha and A.~Zahed, \emph{{Crossing Symmetric Dispersion Relations for Mellin Amplitudes}}, \href{https://doi.org/10.1103/PhysRevLett.126.211602}{\emph{Phys. Rev. Lett.} {\bfseries 126} (2021) 211602} [\href{https://arxiv.org/abs/2101.09017}{{\ttfamily 2101.09017}}].

\bibitem{Costa:2012cb}
M.S.~Costa, V.~Goncalves and J.~Penedones, \emph{{Conformal Regge theory}}, \href{https://doi.org/10.1007/JHEP12(2012)091}{\emph{JHEP} {\bfseries 12} (2012) 091} [\href{https://arxiv.org/abs/1209.4355}{{\ttfamily 1209.4355}}].

\end{thebibliography}\endgroup
\end{document}